%% file: main.tex
\documentclass[12pt, aps, onecolumn, tightenlines, nofootinbib, eqsecnum, prd, nolongbibliography]{revtex4-2}

\usepackage{graphicx}
\usepackage{tikz}

\usepackage[a4paper,centering,bindingoffset=0mm, inner=25mm, outer=25mm, top=26mm,bottom=36mm,heightrounded]{geometry}

\usepackage{amsmath}	
\usepackage{amssymb}
\usepackage{amsfonts}
\usepackage{mathrsfs}
\usepackage{mathtools}
\usepackage{amsthm}
\usepackage{scalerel}

\usepackage{braket}
\usepackage{slashed}

\usepackage{booktabs}
\usepackage{longtable}	
\usepackage{multirow}
\usepackage{diagbox}

\usepackage{epsf}
\usepackage{epsfig}	
\usepackage{epstopdf}
\usepackage[caption=false]{subfig}

\usepackage{verbatim}
\usepackage{textcomp}
\usepackage{enumitem}

\usepackage{layout}

\usepackage{microtype}

\definecolor{unamblue}{cmyk}{1 0.79 0.12 0.59}

\usepackage[]{hyperref}
\hypersetup{
  colorlinks=true,%
  citecolor=unamblue,%
  filecolor=unamblue,%
	 linkcolor=unamblue,%
	 urlcolor=unamblue,
	 bookmarksnumbered=true,     
	 bookmarksopen=true,         
	 bookmarksopenlevel=1,       
	 pdfstartview=Fit,           
	 pdfpagemode=UseOutlines,
	 pdfpagelayout=TwoPageRight
	}

\newcommand{\dd}{\mathrm{d}}

\newcommand{\cF}{\mathcal {F}} 
\newcommand{\cU}{\mathcal {U}} 
\newcommand{\Newton}{\mathrm{Newt}}

\DeclareMathOperator{\inter}{int} 

\newcommand{\sfA}{\mathsf{A}}

\makeatletter
\newcommand{\subalign}[1]{%
	\vcenter{%
		\Let@ \restore@math@cr \default@tag
		\baselineskip\fontdimen10 \scriptfont\tw@
		\advance\baselineskip\fontdimen12 \scriptfont\tw@
		\lineskip\thr@@\fontdimen8 \scriptfont\thr@@
		\lineskiplimit\lineskip
		\ialign{\hfil$\m@th\scriptstyle##$&$\m@th\scriptstyle{}##$\crcr
			#1\crcr
		}%
	}
}
\makeatother

\newcommand{\facetcolor}{blue!40}

\definecolor{unamblue}{cmyk}{1 0.79 0.12 0.59}

\usepackage{graphicx}
\usepackage{tikz}
\usetikzlibrary{shapes,snakes}
\usetikzlibrary{patterns.meta}
\usepackage{slashed}
\usepackage{epstopdf}
\usepackage{verbatim}	
\usepackage{blkarray}
\usepackage{ytableau}
\usepackage[compat=1.1.0]{tikz-feynman}
\tikzfeynmanset{warn luatex=false}

\tikzset{
  massless/.style={line width=0.2ex},
  massive/.style={double=none,double distance=0.12ex,line width=0.12ex}
}

\newcommand{\loopm}{\ell}

\def\Eqn#1{Eq.~\eqref{#1}}

\def\Eqns#1#2{Eqs.~\eqref{#1} and~\eqref{#2}}

\usepackage[]{todonotes}

\begin{document}
\title{Finite Integrals from Feynman Polytopes}
\author{Leonardo de la Cruz, David A. Kosower}
\affiliation{Institut de Physique Th\'eorique,  CEA, CNRS, Universit\'e Paris–Saclay,
 F--91191, Gif-sur-Yvette cedex, France}
\author{Pavel P. Novichkov}
\affiliation{Institut de Physique Th\'eorique,  CEA, CNRS, Universit\'e Paris–Saclay,
 F--91191, Gif-sur-Yvette cedex, France\\
 \textrm{and\/} 
Department of Physics and Astronomy, Ghent University, 
B--9000 Ghent,
Belgium}

\begin{abstract}
 We investigate a geometric approach to determining 
the complete set of numerators giving rise to finite Feynman
integrals.  Our approach proceeds graph by graph, and makes
use of the Newton polytope associated to the integral's
Symanzik polynomials.  It relies on a theorem by Berkesch,
Forsg{\aa}rd, and Passare on the convergence of Euler--Mellin integrals,
which include Feynman integrals.  We conjecture that a necessary
in addition to a sufficient condition is that all parameter-space
monomials lie in the interior of the polytope.  
We present an algorithm for finding all finite numerators
based on this conjecture.  In
a variety of examples, we find agreement between the results
obtained using the geometric approach, and a Landau-analysis
approach developed by Gambuti, Tancredi, and two of the authors.
 \end{abstract}

\maketitle

\section{Introduction}

\def\eps{\epsilon}
Scattering amplitudes are key ingredients in
theoretical predictions of observables in collider
experiments, in new approaches to
gravitational-wave emission, and in other
contexts.  Higher-order scattering amplitudes are built
out of Feynman integrals, and an understanding of
infrared properties is important to the evaluation
of observables.  In dimensional regularization, 
these infrared (and ultraviolet) properties appear through poles in
the dimensional regulator $\eps$. An interesting subclass of Feynman integrals consists of those free of poles.

  Characterizing such finite Feynman integrals is helpful to
improving computations of scattering amplitudes.
Two of the authors, with Gambuti and Tancredi (GKNT)
recently described~\cite{Gambuti:2023eqh}
a systematic method based on the Landau equations~\cite{Bjorken:1959fd,Landau:1959fi,Nakanishi1959}
and on computational algebraic geometry methods to determine
all numerators giving rise to finite Feynman integrals.
The construction of such integrals was previously considered in Refs.~\cite{vonManteuffel:2014qoa, Anastasiou:2018rib} and applied in Refs.~\cite{vonManteuffel:2015gxa, vonManteuffel:2017myy}. An extension of these ideas was presented in Ref.~\cite{Agarwal:2020dye}.

Feynman integrals are  generally divergent. 
In a parametric representation the domain of finiteness
of an integral is characterized in a theorem by Nilsson and
Passare~\cite{2010arXiv1010.5060N},  later generalized by
those authors together with 
Forsg{\aa}rd (BFP)~\cite{2011arXiv1103.6273B}. These theorems treat Euler--Mellin integrals generally, of which Feynman integrals are a subclass.  The key element in these theorems is the Newton polytope associated with the Symanzik polynomials.  This polytope reflects the
geometry of \emph{exponents\/} of Feynman-parameter
monomials, as we explain in Sect.~\ref{polytopes}.  Membership
in the so-called relative interior of the polytope will
determine whether an integral is finite.

These polytopes have been used for studies of the
convergence domains for general Feynman integrals, crucial to
interpreting Feynman integrals as 
$\sfA$-hypergeometric functions~\cite{GELFAND1990255,%
2010arXiv1010.5060N,2016arXiv160504970N,Schultka:2018nrs,%
delaCruz:2019skx,Klausen:2019hrg,Klausen:2021yrt,%
Ananthanarayan:2022ntm,2011arXiv1103.6273B,delaCruz:2024ssb}. 
They also provide a link to tropical 
geometry~\cite{2015Forsgaard,Panzer:2019yxl, Borinsky:2020rqs,%
Arkani-Hamed:2022cqe, Borinsky:2023jdv,Salvatori:2024nva}, and can be
used in the method of regions~\cite{Ananthanarayan:2018tog,%
Gardi:2022khw} and in the evaluation of integrals using 
sector decomposition~\cite{Borowka:2015mxa,Heinrich:2021dbf}.  Recently, Gardi, Herzog, Jones, and Ma (GHJM) \cite{Gardi:2024axt} have studied singularities of a particular class of non-planar integrals using Newton polytopes.

Nontrivial numerators are typically present in some master
integrals beyond one loop.  They are also essential
to constructing finite integrals without introducing 
doubled propagators or shifted dimensions.  Nontrivial
numerator polynomials in the loop momenta and external momenta
lead to nontrivial numerator polynomials in the Feynman
parameters.  The treatment of complete polynomials is beyond
the scope of the BFP theorem, which does however allow
for the treatment of numerator monomials individually.

We conjecture that the monomial integrals can be
analyzed independently to determine the convergence of
the Feynman integral as a whole.  In this paper,
we build an approach based on this conjecture, and
compare the independent numerators yielding finite integrals
with those obtain by the GKNT Landau analysis.  As we
shall see, there are subtleties in the comparison, but
the results support the conjecture and the approach.  
The methods proposed in this paper then provide an
independent way of finding the complete set of finite
Feynman integrals for a given topology.

This paper is organized as follows. In the next section,
we establish notation, describe Newton polytopes, review
the BFP theoreom, and give a simple example of its application.
In Sect.~\ref{algorithm}, we present an algorithm for
constructing numerators yielding finite integrals using
polytopes.  In Sect.~\ref{examples}, we apply the algorithm
to several one- and higher-loop examples, and also discuss
the comparison to the GKNT Landau analysis.
We summarize in Sect.~\ref{the-end}.  Two appendices discuss
technical details, and a third presents arguments backing our
key conjecture.

\def\ellset{{\pmb{\ell}}}
\def\kset{\pmb{k}}
\def\sset{\pmb{s}}
\def\Numer{\mathcal{N}}
\def\Den{\mathcal{D}}
\def\Integ{\hat I}
\section{Feynman integrals with numerators}
\label{core}

\subsection{Notation}
\def\xvec{\pmb{x}}
\def\nvec{{\pmb{n}}}
\def\Pvec{\pmb{P}}
\def\gvec{\pmb{g}}
For polynomials, we employ a multi-index notation, 
\begin{equation}
{\xvec}^{\nvec} \equiv x_1^{n_1} \cdots x_E^{n_E}\, ,
\end{equation}
for the variables $\xvec=(x_1, \dots,x_E)$.
\noindent where $\nvec\in \mathbb{K}^E$. For polynomials
$b_1(\xvec), \dots, b_M(\xvec)$ in the variables $\xvec$, 
the multi-index notation for products of  polynomials reads, 
\begin{equation}
\gvec(\xvec)^{\nvec} \equiv
 g_1(\xvec)^{n_1} g_2(\xvec)^{n_2} \cdots
 g_M(\xvec)^{n_M}\,,  
\end{equation}
where here $\nvec \in \mathbb{K}^M$.  We will typically
have $\mathbb{K}=\mathbb{N}$
or $\mathbb{K}=\mathbb{Z}+\mathbb{Z}\epsilon$,
where $\epsilon$ is the dimensional regulator.

\subsection{Parametric Representation of Tensor Integrals}
We study Feynman integrals with nontrivial 
numerators.  The numerators are polynomials in Lorentz invariants involving loop momenta $\ell_j$ 
and external vectors.
It will be sufficient in general for us to consider 
external momenta $k_j$ for
the set of external vectors, so the numerators
are polynomials in, 
\begin{equation}
    \{\ell_i\cdot\ell_j,\, \ell_i\cdot k_j\}\,.
    \label{BaseVariables}
\end{equation}
We take the coefficients to be rational functions in the kinematic parameters (external invariants, external and internal masses).
In particular, we do not allow the spacetime dimension~\(D\) in the coefficients.
In a slight abuse of language, we will refer to the degree 
of the polynomial in the collection
of loop momenta as the rank.  We write a Feynman integral corresponding to a graph
with $E$ internal edges and $L$ loops as follows,
\begin{equation}
    \Integ[\Numer(\ellset,\kset)] = \int 
      \prod_{j=1}^L \frac{\dd^D\ell_j}{i\pi^{D/2}} \;
      \frac{\Numer(\ellset,\kset)}{\Den_1\cdots \Den_E}\,,
\label{integral-loop-withnumerator}
\end{equation}
where $\ellset$ denotes the $L$ loop momenta, and $\kset$ 
the set
of $n$ independent external momenta. 
The denominators have the form,
\begin{align}
 \Den_e = (M_e)^{jr}\ell_j\cdot \ell_r 
     + 2 (Q_e)^{jr} \ell_j\cdot k_r +J_e +i\varepsilon\,, 
\end{align}
where the matrices $M_e$ and $Q_e$ have dimensions 
$L\times L$ and $L\times n$ respectively.
In order to express the parametric representation we define
the matrices,
\def\Qc{\widetilde Q}
\def\Mc{\widetilde M}
\def\Jc{\widetilde J}
\def\feynp{\alpha}
\def\feynpset{\pmb{\feynp}}
\begin{equation}
\Mc^{jr}= \sum\limits_{e=1}^E \feynp_e M_e^{jr}, 
\quad  \Qc^{j\mu} = 
  \sum\limits_{e=1}^E \feynp_e Q_e^{jr}k_r^\mu\,,
\end{equation} 
and the scalar,
\begin{equation}
\Jc=\sum\limits_{e=1}^E \feynp_e J_e \, ,
\end{equation}
where $\feynp_e$ are the Feynman parameters.
The Symanzik polynomials are then,
\begin{equation}
\mathcal{U} = \det(\Mc), \quad 
\mathcal{F} = \det(\Mc)\left(\Jc
   -\left(\Mc^{-1}\right)^{ij} \Qc^{i}\cdot \Qc^{j}
   \right)/\mu^2\,, 
\end{equation}
where as is conventional, we have scaled $\mathcal{F}$ by
the `renormalization' scale $\mu$ to make it dimensionless
(as needed for the Lee--Pomeransky representation; we will leave
it implicit below).  
The first Symanzik polynomial $\mathcal{U}$ is a homogeneous 
polynomial of degree $L$, while the second Symanzik 
polynomial $\mathcal{F}$ is homogeneous of degree $L+1$.  
In Euclidean kinematics, the Symanzik polynomials 
$\mathcal{U}$ and $\mathcal{F}$ are positive semi-definite functions of the Feynman parameters.
These polynomials can also be obtained from the topology 
of the corresponding graph.  For a review
of their properties and the graph connection, see
Ref.~\cite{Bogner:2010kv} and references therein. 

We are interested in integrals that have not only
a monomial in the numerator $\Numer$ of 
\Eqn{integral-loop-withnumerator}, but a sum
of many monomials.  It is possible to write down
a closed-form parametric expression originating from
a monomial~\cite{Heinrich:2008si}, and then sum over
such expressions.  This approach is somewhat tedious 
to implement.  We can also recast the
calculation of the parametric representation resulting
from a single term in the numerator 
of~\Eqn{integral-loop-withnumerator} as a worldline-like 
integral so we can construct its parametric representation 
recursively via Wick contractions.  
See Appendix \ref{tensor-parameters} for details. 

\def\NumerP{{\widetilde\Numer}}
We can bring the parametric representation of 
 \Eqn{integral-loop-withnumerator} into the 
form,
\begin{align}
\Integ[\Numer(\ellset,\kset)]= 
\Gamma\biggl(E-\left\lfloor\frac r 2\right\rfloor
  -\frac {LD}{2}\biggr)
\int \dd^E \feynp \, 
   \delta\Bigl(1-\sum_{i\in A} \feynp_e\Bigr)
		\cU^{E-D/2(L+1)-r} \cF^{DL/2-E} 
         \NumerP(\feynpset),
\label{general-parametric}
\end{align}
where $r$ is the rank of the highest-rank term in 
$\Numer(\ellset,\kset)$; and where the Cheng-Wu theorem
allows us to choose $A$ to be any nonempty subset of 
$\set{1, \dots, E}$. Here  $\NumerP(\feynpset)$ 
is a polynomial in the Feynman parameters $\feynpset$ with coefficients
that are polynomials in the external invariants $\sset$
and in $D$.  It arises from transforming 
$\Numer(\ellset,\kset)$ along with the denominator
of \Eqn{integral-loop-withnumerator} to a parametric form.
Also, $\lfloor r/2\rfloor$
denotes the nearest integer less or equal to $r$. 

We can decompose the numerator in a basis of monomial 
terms as follows,
\begin{align}
\NumerP(\feynpset) = \sum_i c_i \feynpset^{\nvec_i}\,,
\end{align}	
where $\nvec_i\in \mathbb N^{E}$ and the sum runs over all
monomials in $\NumerP(\feynpset)$.  We may choose  $A=\set{o}$, where $o\in \set{1, \dots, E}$, to obtain,
\begin{equation}
\Integ[\Numer(\ellset,\kset)]=  
\Gamma\biggl(E-\left\lfloor\frac r 2\right\rfloor-\frac {LD}
{2}\biggr)
\int \dd^{E-1}\feynp\;
 \cU^{E-D/2(L+1)-r} \cF^{DL/2-E} \NumerP(\feynpset)\,,
 \label{ParametricIntegral}
\end{equation}
where the measure of the integral is now 
$\dd^{E-1}\feynp=
\dd \feynp_1 \cdots \widehat{\dd {\feynp_o}} 
\cdots \dd \feynp_E$, with the hat indicating the omission of 
the corresponding measure factor, and the integral now taken
over the positive orthant. We will choose $A=\{E\}$ below.
 
We can also obtain the Lee--Pomeransky (LP)
representation of Feynman integrals \cite{Lee:2013hzt}
from \Eqn{general-parametric}.  With unit numerator,
this representation reads,
\begin{equation}
\Integ_{\text{LP}}[1]= \frac{\Gamma(D/2)}{\Gamma((L+1)D/2-E) }
		\int \dd^E \feynp  \frac{1}{ \mathcal {G}^{D/2}} \, ,
\label{Leerep}
\end{equation} 
where $\mathcal G\equiv \mathcal{U}+\mathcal{F}$, and where
the integral is taken over the positive orthant. The LP 
representation for \Eqn{general-parametric} can be found as
follows.  By construction the numerators are homogeneous 
polynomials of degree $rL$. Assuming that 
$E-(D/2)(L+1)-r\ne 0$  the LP representation of the 
integral is given by, 
\begin{equation}
\Integ_{\text{LP}}[\Numer(\ellset,\kset)] = 
\frac{\Gamma (r+D/2) 
\Gamma(E-\left\lfloor\frac r 2\right\rfloor-\frac{LD}{2})}
{\Gamma(r+D/2 (L+1)-E)\Gamma(E-LD/2)}
		\int\dd^E \feynp
		\frac{\NumerP(\feynpset)}{{\mathcal G}^{D/2+r}}\,.  
\end{equation}
One can prove this result by inserting 
$1= \int\dd t \, \delta(t-\sum_e \feynp_e) $ into 
the right-hand size, rescaling the parameters by $t$, and 
integrating over $t$.

\subsection{Polytopes}
\label{polytopes}
Our main computational tool will be polytopes of 
Symanzik polynomials.  We give a very brief overview to the basic
notions; the reader may consult  Chapter~2 of the Sturmfels 
book~\cite{sturmfels:1996} for a pedagogical exposition 
of the necessary concepts.  If $f(\xvec)$ is a polynomial
with coefficients in $\mathbb{K}$, that
is $f(x_1, \dots, x_E) \in \mathbb K[x_1, \dots x_E]$,
we can write it as follows,
\def\mvec{{\pmb{m}}}
\begin{equation}
f(\xvec)=\sum_{\pmb{m} \in B} c_{\pmb{m}} \xvec^{\pmb{m}}\,,
\end{equation} 
where $B$ is a (finite) subset of $\mathbb Z_{\ge0}^E$. 
We will be focused on the geometry of \emph{exponents\/},
not of coefficients.
The support of $f$ is then,  
\def\supp{\mathop{\rm supp}\nolimits}
\def\conv{\mathop{\rm conv}\nolimits}
\begin{equation}
\supp(f) \equiv \set{\mvec | c_\mvec\ne0} \,. 
\end{equation}
We will be interested in the \textit{convex hull\/} of
this set, that is, the set of all points which can be represented as certain  linear combinations of the elements in $\supp(f)$.
This convex hull is called the Newton polytope, denoted
by $\Newton(f)$,
\begin{equation}
\begin{aligned}
\Newton (f) &=
\conv(\supp(f)) \equiv
\\&
 \left\{ \sum_{i=1}^{|B|}\lambda_i \mvec_i \,\bigg|\,
   \lambda_1, \dots, \lambda_{|B|}\in  \mathbb R_{>0}, 
   \sum_{i=1}^{{|B|}}\lambda_i=1,
   \mvec_i \in \supp(f)\right\} .
\end{aligned}
\label{newton-1}
\end{equation}
Its dimension is the dimension of the affine space
(lines, planes, or higher-dimensional planes in
$\mathbb{R}^E$ not necessarily passing through the origin)
which it spans.  A polytope is full dimensional if
its dimension is the same as the dimension of the
exponent vectors $\mvec$.  (The dimension can be smaller.)
 Any convex polytope can be described
as the convex hull of a finite set of extreme points or
alternatively as the convex hull of the intersection of a 
finite number of half spaces. These are known as the vertex 
($V$-) representation and the half-space 
($H$-) representation respectively.  
While \Eqn{newton-1} gives a finite
representation of the Newton polytope of $f$, it is in 
general redundant. It contains more vertices than needed to 
describe the polytope: some $\mvec_i$ will be in the 
interior.  The $V$-representation removes these superfluous 
vertices. The $H$-representation is given by $M$ 
inequalities, 
\begin{equation}
 P=\set{\mvec \in \mathbb{Z}^E | A\mvec-b\ge 0}\, ,
 \label{HRepresentation}
\end{equation}  
where $A$ is an $M\times E$ matrix and $b$ is
an $E$-vector.  We can define a face of a polytope $P$
as in Ref.~\cite{sturmfels:1996},
\def\face{\mathop{\rm face}\nolimits}
\begin{equation}
\face_w (P) \equiv \{ \mvec \in P | 
 w \cdot \mvec \ge w \cdot v \quad \text{for all } v\in P\},
\end{equation}
\begin{figure}[htb]
	\centering
		\input{faces_Example.tex}
		\caption{Polytope with 9 faces. The relative interior of this polytope is the shaded area with the edges and vertices removed. }
   \label{fig:example-faces}
\end{figure}
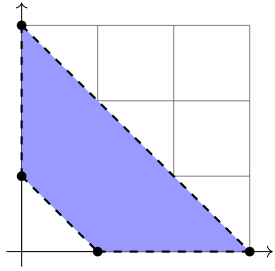
\noindent  for a given $w\in \mathbb R^E$.	 A finite set of $w$s
suffices to enumerate all distinct faces.
Every subset $F$ of $P$ given by some $w$ is called a face 
of $P$. For example, the polytope in 
Fig.~\ref{fig:example-faces} has nine faces, 
of dimensions ranging from 0 to 2.  This includes the
polytope itself.  
The polytope itself is called an improper face, while the
other faces are called proper faces.   
The \textit{relative interior\/} of a polytope $P$ is 
the polytope with its proper faces removed \cite{ziegler1994}. The relative interior is nontrival even for polytopes that are not full dimensional, whereas the interior would be empty in this case. We will primarily use the relative interior in computations instead of the interior because that is what available codes yield. For our purposes (full dimensional polytopes)  the interior and the relative interior are the same.

The Minkowski sum of two 
convex\footnote{Newton polytopes are by construction convex
polytopes.} polytopes $P, Q \in \mathbb R^E$ is given
by a sum of their points,
\begin{equation}
    P+Q \equiv \set{p+q|p\in P, q\in Q} \subset \mathbb R^E\,.
\end{equation}
The sum of two or more polytopes is also a 
polytope. See Chapter~3 in Ref.~\cite{schneider_2013} for a 
detailed discussion of Minkowski addition and subtraction. 
For Newton polytopes, we will also make use of scalar 
multiplication, 
\begin{equation}
\lambda P \equiv \set{\lambda p| p\in P}\,,\quad 
\lambda \in \mathbb R \, .
\end{equation}
For two polynomials $f,g$, we have the result,
\begin{equation}
  \Newton (f g)=\Newton(f)+\Newton(g) \,;
  \label{Minkowski-sum}
\end{equation}
and for a (positive) power $n$ of a polynomial, the result,
\begin{equation}
  \Newton (f^n)= n\,\Newton(f) \, .
  \label{Minkowski-scaling}
\end{equation}

\def\FeynmanPolytope{P_F}
\def\relint{\mathop{\rm relint}\nolimits}
 \subsection{Convergence of tensor integrals}
\label{convergence}
The convergence of Feynman integrals appearing in \Eqn{ParametricIntegral}, where $\NumerP$ is
a lone monomial,
is determined by a theorem due to 
Berkesch, Forsg{\aa}rd and Passare (BFP) \cite{2011arXiv1103.6273B}.   The theorem was
strengthened by Schultka~\cite{Schultka:2018nrs}.
 The theorem looks
at the Newton polytope of the integrand, with the
measure absorbing factors of $1/\alpha_e$ to make
it projective.  That is, we are interested in the
Newton polytope,
\begin{equation}
\Newton\bigl(
  \bigl[\cU^{E-D/2(L+1)-r} \cF^{DL/2-E}\bigr]^{-1}\bigr)\,.
\end{equation}
Defining,
\begin{equation}
    \begin{aligned}
    n_\cU &\equiv r-E+D/2(L+1)\,, \\
    n_\cF &\equiv E-DL/2\,,    
    \end{aligned}
\end{equation}
 and using \Eqns{Minkowski-sum}{Minkowski-scaling}, we can
 reexpress this Feynman polytope as a weighted sum,
 \begin{equation}
 \FeynmanPolytope =
     n_\cU\,\Newton(\cU) + n_\cF\,\Newton(\cF)\,.
 \end{equation}
We must require that $\cU$ and $\cF$ be 
completely non-vanishing
in the following sense.   The notion relies on a
\textit{truncation\/} $g_F$ of a polynomial $g$ to a face $F$,
which is the sum of those monomials in $g$ whose
exponent vectors lie on the face $F$.
A polynomial $g$ is completely non-vanishing 
on $\mathbb R_{> 0}^E$ if for each face $F$ of 
$\FeynmanPolytope$ the 
truncated polynomial $g_F$ has no zeros on 
$\mathbb 
R_{> 0}^E$. This property is trivially satisfied if the coefficients in the polynomials have all of the same sign. This is always the case for $\cU$ and 
 it is also the case of the second 
Symanzik polynomial $\cF$ of planar integrals, where a Euclidean 
region exists.

For full-dimensional $P_F$ and
for completely non-vanishing polynomials $\cU$ and $\cF$,
and $n_\cU\ge 0, n_\cF>0$,
the BFP theorem tells us that 
the integral converges and defines an analytic function in
the Mandelstam invariants and in the domain for
the exponents,
\begin{equation}
\set{ \mvec \in \mathbb{C}^E|
\mvec+\pmb{1}\in \inter(\FeynmanPolytope)
		} \,,
\label{BFP}
\end{equation}
where $\inter{P}$ denotes the interior of
$P$. As explained in the previous 
subsection\footnote{When the Feynman polytope is
less than full dimensional, the integral is expected
to diverge; indeed, such integrals are known to
not be absolutely convergent~\cite{Schultka:2018nrs}.} we can use the relative interior of $P$ instead of the interior, writing the domain of the exponents as,
\begin{equation}
\set{ \mvec \in \mathbb{C}^E|
\mvec+\pmb{1}\in \relint(\FeynmanPolytope)
		} \,,
\label{BFPrelint}
\end{equation}
where $\relint(\FeynmanPolytope)$ denotes the relative interior of $P$.
In the special case when $n_\cU=0$, that is, 
$r=E-\frac D2 (L+1)$, convergence is determined only 
by $(E- DL/2)\Newton(\mathcal F)$. 
When $n_\cU < 0$, we similarly use the latter polytope and treat $\cU$ as part of the numerator.

A similar
application of the theorem can be stated for the 
integral in the LP representation using the
modified Newton polytope,
\begin{equation}
P_{\text{LP}}=\left(r+\frac D 2 \right) \Newton (\mathcal G) \, .
\end{equation}
Notice that for rank $0$, $\mvec=0$ since there are no monomials produced in \Eqn{general-parametric} and thus the integral
will be finite if $\FeynmanPolytope$ contains $\pmb{1}$
in its interior.
Integrands for full-dimensional
polytopes with completely non-vanishing
polynomials that contain exponents lying 
outside the domain of convergence are known
to diverge~\cite{Schultka:2018nrs}.  The theorems are silent on
the fate of integrands failing the complete non-vanishing
condition, as is the case of nonplanar integrals.
We will see in examples that even upon relaxing this
condition, the BFP convergence domain is correct.

Now consider parametric Feynman integrals where $\NumerP$ is a sum of several monomials.
It is clear that if the integral for each monomial taken separately is finite, then the integral of the complete numerator is finite.
Our conjecture is that this is also a necessary condition; the integral of the complete numerator will be finite \emph{only\/} if the integral of each monomial separately is finite.
That is, we can analyze parameter-space numerators monomial by monomial.
See Appendix~\ref{app:several-mons} for an argument supporting this conjecture.
 
\subsection{A Triangle Example}
\def\tri{\triangle}
\begin{figure}[htb]
\centering
\begin{tikzpicture}
  \begin{scope}[massless]
  \draw(0,0.732) -- node[left] {$1$} (-1, -1) -- node[below] {$3$} (1, -1) -- node[right] {$2$} cycle;
  \draw(0,0.732)--+(90:0.8) node[left] {$k_2$};
  \end{scope}
  \pgfsetblendmode{multiply} 
  \begin{scope}[transparency group]
  \draw[massive] ($(-1,-1)-(-150:0.04)$)--+(-150:0.84) node[left] {$k_1$};
  \draw[massive] ($(1,-1)-(-30:0.04)$)--+(-30:0.84) node[right] {$k_3$};
  \end{scope}
\end{tikzpicture}
\caption{Triangle with massless internal lines and 
two massive external legs}
\label{toy-triangle}
\end{figure}
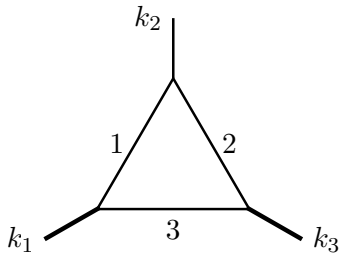

Before examining more complicated integrals, let us apply  
the BFP theorem via \Eqn{BFP} to the triangle integral, 
shown in Fig.~\ref{toy-triangle},
\begin{equation}
\hat I_{\tri}=\int\frac{\dd^D\loopm}{i \pi^{D/2}} 
\frac{\Numer(\ell,\kset)}{\Den_1 \Den_2 \Den_3}\,,
\end{equation}
where $\Den_1=\ell^2$, $\Den_2=(\ell-k_2)^2$, and 
$\Den_3=(\ell+k_1)^2$.  After conversion to parametric
form, the integral can be expressed as a linear combination 
of monomials whose set of exponents we denote by $B$,
\begin{equation}
		\Integ_{\tri,r}= \sum_{\mvec \in B} c_\mvec (\sset) 
  \int \frac{\dd^3 \feynp}{\feynp_1\feynp_2\feynp_3}\;
  \delta(1-\feynp_1-\feynp_2-\feynp_3)  \feynp^{\mvec+1} \, 
  \mathcal U^{3-D-r} \mathcal F^{D/2-3}\,,
  \label{toy}
\end{equation}
where $\cU=\feynp_1+\feynp_2+\feynp_3$ and 
$\cF=-(k_1^2 \feynp_1\feynp_3+k_3^2 \feynp_2\feynp_3)$. 
We choose the set $A=\set{3}$ here, setting $\feynp_3$ 
to $1$. In Euclidean kinematics, $\cF$ and $\cU$ are positive for $\feynp_{1,2} \in \mathbb R_{\ge0}^2$. In the limit $D=4$,
$n_\cU$ and $n_\cF$ are both positive, sufficient
for application of the BFP theorem. 
The convergence of the integral is determined by
the Newton polytope,
\begin{equation}
P_\tri = (r+D-3)\,\Newton(\cU)+(3-D/2)\, \Newton(\cF)\, .
\end{equation}
\begin{figure}[htb]
   \centering
   \subfloat[$r=0$]{%
     \label{fig:grappoltriangle}
     \input{triangle_poly1.tex}
   }
   \qquad
   \subfloat[$r=1$]{%
     \label{fig:grappoltriangle2}
     \input{triangle_poly2.tex}
   }
   \qquad
   \subfloat[$r=2$]{%
     \label{fig:grappoltriangle3}
     \input{triangle_poly3.tex}
   }
\caption{Weighted polytopes for the two-mass triangle
for different numerator ranks. 
There are no lattice points in the relative interior for $r=0$, and one and three for $r=1$ and $r=2$, respectively.}
\label{fig:grappoltriangles}
\end{figure}
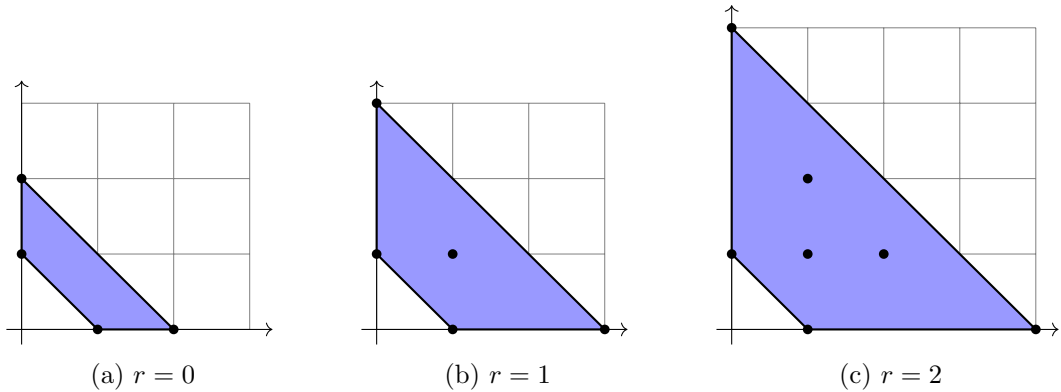
We are interested in all points that lie in the relative 
interior of $P_3$, in particular the integer lattice
points corresponding to monomials in \Eqn{toy}.  
We consider integer values for 
$\mvec \in \mathbb Z^2$ and also take $D\rightarrow 4$. 
We need to determine what vectors lie in the relative 
interior of $P_3$ as the rank increases (see 
Fig.~\ref{fig:grappoltriangles}). There are no interior 
lattice points for $r=0$; the integral $\hat I_{\triangle,0}$ is
accordingly divergent. There is a single point in the 
relative interior for $r=1$, namely $\mvec +\pmb{1} =(1,1)$, and 
three for $r=2$.  

The lone relative-interior point at $r=1$ implies that
$ \hat I_{\triangle,1}$ with $\pmb{m}=\pmb{0}$ is finite.  To find the corresponding
numerator in the original loop-momentum representation,
assume $k_1\ne k_3$, and consider the ansatz,
\begin{equation}
   \label{ansatz-triangle}
   \Numer(\loopm,\kset) = 
       c_1 \loopm\cdot k_1+c_2 \loopm\cdot k_2 \,
\end{equation}
for some unknown coefficients $c_1, c_2$.
Converting to the parametric form and matching coefficients
gives us $c_1=0$, so that the finite rank-1 numerator is 
simply $\Numer=\loopm\cdot k_2$. 

It is also instructive to compare with the calculation
for the LP representation. The Newton polytope now reads,
\begin{equation}
   P_{\tri,\text{LP}}=(r+2)\,\Newton ({\mathcal G})\,.
   \end{equation}
This polytope lives in $\mathbb R^3$ and has a single point in its interior, given by $\mvec+\pmb{1}=(1,1,2)$.
This corresponds to the exponent vector $(0,0,1)$. 
For the ansatz~\eqref{ansatz-triangle}, the parametric 
representation of the numerator is,
\begin{equation}
   \NumerP(\feynpset,k_1^2,k_3^2) = 
    \feynp_3\left[
       \frac{1}{2} c_2 \left(k_1^2-k_3^2\right)
        -c_1 k_1^2\right]
      +\feynp_2\left[
       \frac{1}{2} c_1 \left(k_3^2-k_1^2\right)\right]\,.
\end{equation}
As the only finite monomial is $\feynp_3$, the coefficient 
of $\feynp_2$ must vanish.  This leads to the same 
result for the numerator as above.  In this simple example, 
the number of the interior points matches in both 
representations; this will not be the case in general. 
	
Both representations can be used interchangeably, except 
when the LP representation is singular as mentioned earlier. 
Our goal is to systematize this approach to general 
integrals and numerators of any rank.   We will use
the standard Symanzik representation.
\section{Finite numerators from  polytopes}
\label{algorithm}
We now present an algorithm for constructing finite
numerators of a given rank, generalizing the triangle
example. Suppose we are interested in determining the 
convergence of a general Feynman integral with a numerator 
of (maximum) rank $r$.  \Eqn{BFP} instructs us to build
the Newton polytope $P_F$ for the integral, take 
all possible exponents $\mvec$,
and determine whether each corresponding point 
$ (\mvec+\pmb{1})\in \relint(P_F)$. 
We could do this systematically by constructing the 
$H$-representation of $P_F$ and testing each point
in turn to see whether it satisfies all inequalities in
\Eqn{HRepresentation}.  This would be slow (because it would be overkill), 
though
a variant of this idea is feasible, as discussed below.

Alternatively, we can recast the determination
as a linear programming optimization 
problem~\cite{fukudafrequently}.   
\def\relintpts{L}
Denote the set of \emph{lattice} 
relative-interior points at each rank by $R_{F,r}$. Finding this set for a general polytope is a very 
difficult computational problem. 
Several packages in established computer-algebra systems
have implementations of algorithms for this purpose,
and there are also more specialized packages 
such as \textsf{NConvex}~\cite{NConvex2022.09-01} in 
GAP~4~\cite{GAP4}.  We will call this approach
to obtaining the set of relative interior points, together
with remaining steps detailed below, Algorithm~N. Finding the $H$-representation is fast with  
\textsf{NConvex} but obtaining the total number of lattice points or  the set of interior points is generally slower.

Solving the general problem is 
overkill for the
the Newton polytopes arising from Feynman integrals. 
We can proceed alternatively using a conjecture for the generating
function of Feynman polytopes.
This approach is modeled on that used in Barvinok's 
algorithm~\cite{Barvinok94} for counting lattice points in a polytope 
efficiently.  This algorithm is implemented in 
\textsf{LattE}~\cite{DELOERA20041273}.   The algorithm
constructs a generating function for a polytope $P$,
which can also be expressed in the form,
\begin{equation}
f_P(\xvec)=\sum_{\mvec \in P \cap \mathbb Z^E } 
  \xvec^\mvec \,.
\end{equation}
Barvinok obtains the number of lattice points by
evaluating $f_P(1)$.  With the analytic form of
the generating function, we can scan all monomials,
and determine the relative interior points using
the matrices $A$ and $b$ of the 
$H$-representation~\eqref{HRepresentation} of $P$.
The relative interior points are those exponents
where all inequalities hold strictly. 

\def\Unit{\mathop{\rm Unit}\nolimits}
For a Feynman polytope $P_F$, we conjecture that 
the generating function is given by,
\begin{equation} 
   f_{P_F}(x) = 
   \Unit\bigl(\cU^{(r+D/2(L+1)-E)}\cF^{E-DL/2}\big|_{D=4}
   \bigr)\, , 
\label{gen-function}
\end{equation}
where $\Unit$ stands for the operation of expanding
its argument, and setting coefficients of all monomials
to $1$.  (Its argument is a polynomial here, as $r$ is
a positive integer.)
Using \Eqns{Minkowski-sum}{Minkowski-scaling}, we can see
the Newton polytope of $f$ is precisely $P_F$.  
We will call this approach to obtaining the relative
interior points, together with the remaining steps below,
Algorithm~G. 

The idea is simple. We classify the points produced by the 
generating function \eqref{gen-function} as points either in the 
relative interior or in the proper faces of the polytope. A point in a face satisfies at least one equality of \Eqn{HRepresentation}. We simply exclude those points  from the interior-point counting. In the Euclidean case the coefficients of the Symanzik polynomials are positive so there cannot be cancellations. These cancellations can occur in general but in all our examples we found that the counting based on Algorithm G and N match. Newton polytopes that satisfy the property that each monomial is a lattice point are said to be \emph{saturated} \cite{monical2019newton}. It would be interesting to prove this property in general.

While testing all possible interior points would be
computationally expensive, we may note that not every
point in the Feynman polytope can actually be obtained
from the conversion to parametric form of a loop-momentum
numerator.  (In fact, only a small fraction are.)
We can build the set of candidate relative-interior
points by scanning all possible monomials in the 
loop momentum representation; converting an integrand
consisting solely of a given monomial to the 
Feynman-parameter representation; factoring
out a power of $\cU$ and $\cF$ common to all
loop-momentum monomials; and recording all
multi-exponents of the Feynman-parameter monomials 
that arise in the candidate set.  We then test each
of the candidates to see whether it satisfies (strictly)
all the inequalities in \Eqn{HRepresentation}.  
We will call this approach, together with the remaining 
steps, Algorithm~L.

\def\monomset{W}
\def\genericMonom{w}
Given the set of interior points obtained
by following the initial steps of any of
Algorithms~N, G, or L, we derive
constraints on the form of possible numerators.
Start with all independent Lorentz 
invariants~\eqref{BaseVariables} as base variables,
and build the set $\monomset$ of all monomials $w$ in these variables
up to the desired rank $r$ in the loop momenta.  Include
a monomial at rank $0$.  We write an ansatz,
\begin{equation}
    \Numer(\ellset,\kset) = 
     \sum_{\genericMonom\in \monomset} 
        c_{\genericMonom}\, \genericMonom\,;
        \label{gen-ansatz-numerator}
\end{equation}
the coefficients $c_{\genericMonom}$ are dimensionful so that
all terms in $\Numer$ are of the same engineering dimension.
We can convert the integral with this numerator to 
parametric form and factor out a uniform power of
$\cU$ and $\cF$, yielding a numerator polynomial
in the Feynman parameters,
\begin{equation}
    \NumerP(\feynpset,\sset) =
    \sum_{\mvec\in C} a_\mvec(c,\sset,D)\,\feynpset^\mvec\,,
\end{equation}
where the set of exponents $C$ covers all monomials
emerging from $\Numer$, and where the coefficients $a$
in this representation depend linearly on 
the $c_{\genericMonom}$,
and polynomially on the space-time dimension $D$.
By construction, the degree of $\NumerP$ is $rL$.
We obtain the most general finite numerator by
setting to zero the coefficients of all points not in
the relative interior, that is finding the solutions
to, 
\begin{equation}
    \set{a_\mvec = 0 \,|\, \mvec \in \supp(\NumerP) 
    \backslash R_{F,r}}\,.
    \label{InteriorEquations}
\end{equation}
In imposing the vanishing of these $a_\mvec$, we
require the coefficient of each order in $D$ to vanish
independently, in order to avoid introducing a 
$D$ dependence (and thereby an  $\epsilon$ regulator dependence)
into the coefficients of finite numerators.
In general, this system of equations is overdetermined, 
so we may 
obtain only the trivial solution where 
all $c_{\genericMonom}$ vanish. 
This would be the case,
for instance, if there are no relative-interior 
lattice points in $P_F$.  We can summarize this procedure 
as follows,
\subsubsection*{Finite Numerators Algorithm (FNA)}
\label{finnum-algorithm}
\noindent Input: List of denominators of  Feynman integrals $\Den_e$, desired rank $r$, the number of external momenta, number of loops $L$ and kinematic constraints.  

\noindent Output: Basis for finite polynomials in loop momenta.
\begin{enumerate}
   \item Form the numerator ansatz~\eqref{gen-ansatz-numerator} for the desired rank,
loop and external momenta.  
   \item Compute the parametric representation of the Feynman 
integral with the numerator ansatz, yielding the Symanzik polynomials~$\cU$ and $\cF$, their exponents~$n_\cU$ and $n_\cF$, and the numerator~$\NumerP$.
   \item If $n_\cU \leq 0$, set $P_F=n_\cF \, \Newton(\mathcal F)$; if $n_\cU < 0$, treat $\cU$ as part of the numerator.
\hfil\break 
Otherwise [if $n_\cU > 0$], set $P_F=n_\cU \, \Newton(\mathcal U)
+ n_\cF \, \Newton(\mathcal F)$.  
 \item Check that the Newton polytope is full dimensional (this can be done with \textsf{NConvex} using the command \texttt{IsFullDimensional}).
 
\item Determine the relative interior 
lattice points $\relint(P_F)$  of the Feynman polytope
$P_F$, using one of the approaches N, G, or L
 described
above. [Algorithm L computes a sufficient subset.]

\item Construct the set of monomials corresponding 
to $\relint(P_F)$: \hfill\break
$R_F=\set{\feynp^\mvec\,|\,\mvec \in \relint(P_F)}$. With algorithm L, steps 4 and 5 can be merged.
\item Solve the system of 
equations \eqref{InteriorEquations}. 
\item Substitute the solutions into the ansatz.  
\item A basis of finite numerators is given by the 
coefficients of the surviving free 
parameters $c_{\genericMonom}$ in the ansatz after 
substituting in the solutions.
\end{enumerate}

The BFP theorem implies that any linear combintation
of these numerators yields a finite Feynman integral.
As we shall see, the notion of finiteness here does
not precisely match the GKNT notion of strongly UV locally
finite integrals described in Ref.~\cite{Gambuti:2023eqh}.
We discuss the notion of local finiteness in greater
detail in App.~\ref{appendix-local-finiteness}.
We can of course impose additional constraints on the
basis of finite numerators obtained here.
For instance, we can require strong UV convergence, that
is, that the numerator be free of UV divergences by
power counting.  To compare with the results of
Ref.~\cite{Gambuti:2023eqh}, we will need to impose this
constraint, and also understand how integrals can be
finite in $D=4$ without being locally finite.  
Our approach here is not suitable for discovering 
evanescent or evanescently finite integrals, as it
requires setting $D$ to 4.

\section{Examples}	
\label{examples}
Our goal is to establish a connection between the numerators computed from the GKNT Landau analysis and the numerators computed from our Algorithm. 
The two approaches are quite different, and we may expect the comparison between them to be non-trivial.
Indeed, as we will see the comparison requires to take into account IBP identities. The goal of the following examples is to show the salient features required to compare both approaches.

\subsection{Massless box}
\begin{figure}[htb]
   \centering
   \begin{tikzpicture}
   \begin{scope}[massless]
      \draw(-1,-1)-- +(-135:0.8) node[left] {$k_1$};
      \draw(1,-1)--+(-45:0.8) node[right] {$k_4$};
      \draw(1,1)--+(45:0.8) node[right] {$k_3$};
      \draw(-1,1)--+(135:0.8) node[left] {$k_2$};
      \draw
         (-1,-1)-- node[left] {$2$} (-1.0,1.0)-- node[above] {$3$} (1.0,1.0)-- node[right] {$4$} (1.0,-1.0)-- node[below] {$1$} cycle;
   \end{scope}
   \end{tikzpicture}
\caption{Box integral with all internal and external 
lines massless}
\label{os-massless-box}
\end{figure}
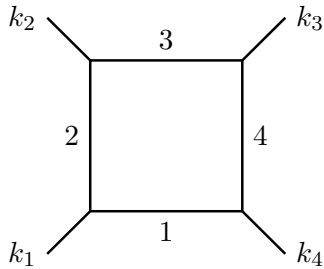
\def\onebox{\square}
As our first example, consider the massless box integrals,
shown in Fig.~\ref{os-massless-box}. 
The inverse propagators are,
\begin{equation}
   \Den_1= \loopm^2, \qquad
   \Den_2=(\loopm-k_1)^2,\qquad
   \Den_3= (\loopm-k_1-k_2)^2,\qquad 
   \Den_4=(\loopm-k_1-k_2-k_3)^2, 
\end{equation}
where $k_i^2=0$, $i=1,\dots,4$. The Symanzik polynomials 
are given by,
\begin{equation}
   \cU_{_\onebox} = 
     \feynp_1+\feynp_ 2+\feynp_ 3+\feynp_ 4,\qquad 
   \cF_{_\onebox}= 
    -(s \feynp_ 1 \feynp_ 3+t \feynp_ 2 \feynp_ 4)\,,
\end{equation}
where $s=(k_1+k_2)^2$ and $t=(k_2+k_3)^2$ are the usual
Mandelstam invariants.  In Euclidean kinematics, they are
both negative.
We need to compute the relative-interior lattice points of
the associated Feynman polytope (or the one associated 
to the LP representation), 
\begin{equation}
   P_\onebox =(4-D/2)\, \Newton(\cF_\onebox)
         +(D+r-4)\, \Newton(\cU_{\onebox})\, .
\label{polytopes-box}    
\end{equation}
The scalar integral corresponds to rank $r=0$.
Setting $D=4$ and $r=0$, we find the
polytope to be a line in $\mathbb R^3$, whose
$V$-representation is,
\begin{equation}
    P_{\onebox,r=0} = \conv(\set{(2,0,2),(0,2,0)})\,.
\end{equation}
(Recall that we have set $\feynp_4 = 1$.)
The polytope here is one-dimensional, but the exponent
vectors are three-dimensional.  According to
Schultka's specialization of the BFP theorem, the integral
is thus divergent.  This agrees with direct calculation,
and with a Landau-based analysis~\cite{Gambuti:2023eqh}.  

The next rank has polynomials linear in the loop momentum
$\ell$.  The matrix $A$ and the vector $b$ for the
$H$-representation~\eqref{HRepresentation} are given
by,
\begin{equation}
 A_\onebox^T=\left(
   \begin{matrix}
      0 & 1 & 0 & 0 & 1 & -1 & 0 & -1 \\
      1 & 1 & 0 & 1 & 0 & -1 & -1 & -1 \\
      1 & 0 & 1 & 0 & 0 & 0 & -1 & -1 \\
   \end{matrix}
   \right), \quad 
   -b_{\onebox,1}^T=\left(
   \begin{matrix}
      -2 & -2 & 0 & 0 & 0 & 3 & 3 & 5 \\
   \end{matrix}
   \right) \,.
\end{equation}
(The minus sign on the left-hand side of 
the second equation matches the
conventions of \textsf{NConvex}.)
At higher ranks, the matrix $A$ is the same,  
but $b$ is instead given by,
\begin{equation}
-b_{\onebox,r}^T=\left(
   \begin{matrix}
      -2 & -2 & 0 & 0 & 0 & 2+r & 2+r & 4+r \\
   \end{matrix}
   \right) \, .
\end{equation}
As the box integral is planar, the Symanzik polynomials
are completely nonvanishing on the faces of the Feynman
polytope.  We can proceed directly to computing the 
relative interior for the rank-one Feynman polytope,
\begin{equation}
    P_{\onebox,r=1} =  2 \Newton(\cF_\onebox)
         +\Newton(\cU_{\onebox})\, ,
\end{equation}
using \textsf{NConvex}.  We find that there 
are no relative interior points.  We can also obtain
the full set of lattice points with \textsf{NConvex}, 
finding 12,
\begin{equation}
\begin{aligned}
\{&(0, 2, 0), (0, 2, 1), (0, 3, 0), (1, 1, 1), (1, 1, 2), 
(1, 2, 0), \\&
(1, 2, 1), (2, 0, 2), (2, 0, 3), (2, 1, 1), (2, 1, 2), 
(3, 0, 2)\}\,.
\end{aligned}
\end{equation}
Alternatively, consider the generating 
function~\eqref{gen-function},
\begin{equation}
f_{\onebox,r=1}=
\Unit(\cU_\onebox\cF^2_\onebox\big|_{\feynp_4=1})
\,.    
\end{equation}
This gives the same 12 lattice points. 
(Even in this simple example, the generating function
method is over three orders of magnitude faster.)
Using the inequalities~\eqref{HRepresentation},
we again find that there are no relative interior lattice points.  
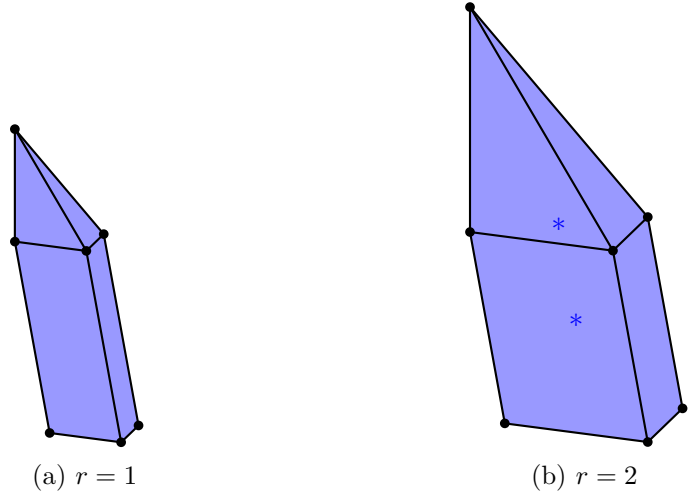
\begin{figure}[tbh]
  \centering
  \subfloat[$r=1$]{%
    \input{rank1-box.tex}    
  }
  \hspace{10em}
  \subfloat[$r=2$]{%
    \input{rank2-box.tex}
  }
\caption{Newton polytopes for the one-loop box at ranks one and two. Stars indicate the relative-interior points.}
\label{fig:OneloopRank2}
\end{figure}
Starting at rank two, whose Feynman polytope is shown
in Fig.~\ref{fig:OneloopRank2}, we do find relative 
interior lattice
points.  The polytope at rank two is,
\begin{equation}
    P_{\onebox,r=2} =  2\, \Newton(\cF_\onebox)
         +2\,\Newton(\cU_{\onebox})\, ,
\end{equation}
The generating-function method again agrees
with \textsf{NConvex}.  At rank two, we find two 
relative-interior points;
and at rank three, eight points.  
At rank two, we have a total of 28 points in the polytope, and
at rank three, 52 points.
With four external
legs, we have three independent external momenta, and
so the set of independent loop-momentum monomials is,
\begin{equation}
\begin{aligned}
\monomset=& \{1, \loopm\cdot k_1, \loopm\cdot k_2,
\loopm\cdot k_3, \loopm^2,
\left(\loopm\cdot k_1\right)^2,
\loopm\cdot k_1\, \loopm\cdot k_2, 
\left(\loopm\cdot k_2\right)^2,
\loopm\cdot k_1\, \loopm\cdot k_3, \\
	&\hphantom{\{} \loopm\cdot k_2\, \loopm\cdot k_3,\left(\loopm\cdot k_3\right)^2\}\,.
\end{aligned}
\end{equation}
Accordingly the numerator ansatz has the form, 
\begin{equation}
\Numer(\loopm,\kset)=c_0+c_{1} \loopm\cdot k_1+ \dots,
\label{RankTwoAnsatz}
\end{equation}
where the ellipsis denotes the remaining 
elements in $\monomset$.

Not every point in the polytope can arise
from the integrand of a Feynman integral.  If we take
the general ansatz~\eqref{RankTwoAnsatz}, and convert
the integrand to parametric form, we find
the set of Feynman-parameter exponents that arises
is a proper subset of the 28 points in the polytope
(after shifting back the latter by the vector $\pmb{1}$),
\begin{equation}
\begin{aligned}
C_{\onebox,2}=\bigl\{&
(0, 0, 0), (0, 0, 1), (0, 0, 2), (0, 1, 0), (0, 1, 1), \\
&(0, 2, 0), (1, 0, 0), (1, 0, 1), (1, 1, 0), (2, 0, 0)
\bigr\}\,.
\end{aligned}
\end{equation} 
This set includes two elements corresponding
to the two relative-interior points,
\begin{equation}
    R_{\onebox,2}=\set{(1,0,1),(0,1,0)}\,,
\end{equation}
as well as exponents corresponding to points
not in the relative interior,
\begin{equation}
B_{\onebox,2}=\set{(0, 0, 0), (0, 0, 1), (0, 0, 2),  (0, 1, 1), 
(0, 2, 0), (1, 0, 0), (1, 1, 0), (2, 0, 0)}\,.
\end{equation}

\def\evec{{\pmb{e}}}
The Feynman integral of $\Numer(\loopm,\kset)$ can be, 
\begin{equation}
\hat I[\Numer(\loopm,\kset)]=
\Gamma\biggl[3-\frac D 2\biggr] 
\sum_{\genericMonom\in \monomset_\onebox}
\sum_{\evec\in C_{\onebox,2}} 
c_\genericMonom f_{w\evec}
  \int \frac{\dd^3 \feynp}{\feynp_1\feynp_2\feynp_3}
     \; \cU_{\onebox}^{2-D} 
        \cF_{\onebox}^{D/2-4}
	\,\feynpset^{\evec+ \pmb{1}}  \, ,
\end{equation}
where $f_{w\evec}$ is a function that depends on the invariants $\pmb{s}$ and $D$.
\def\eqD{\stackrel{\mathclap{D}}{=}}
The coefficient of each Feynman-parameter monomial is
given by,
\begin{equation}
    a_\evec = \sum_{\genericMonom\in \monomset_\onebox} c_w
    f_{\genericMonom\evec}\,.
\end{equation}
We must set the coefficients for $B_{\onebox,2}$ 
to zero, order by order in $D$.
This gives us the following system of equations,
\begin{equation}
	a_{2,0,0}\eqD
	a_{1,1,0}\eqD
	a_{1,0,0}\eqD
	a_{0,2,0}\eqD
 	a_{0,1,1}\eqD
 	a_{0,0,2}\eqD
	a_{0,0,1}\eqD
	a_{0,0,0}\eqD 0 \, .
\end{equation}
where the notation $\eqD$ means each
order in $D$ is equated separately.  This gives us
a total of 16 equations, only eight of which are independent.

The solution to this system is parametrized by three free
coefficients.
A general finite polynomial of rank two is given by a linear
combination of three numerators,
\begin{equation}
\begin{aligned}
\Numer_1 &= 
(2 s+t) k_1\cdot\ell+(s+t) k_2\cdot\ell+s k_3\cdot\ell
- (s+t) \ell^2\,,\\
\Numer_2 &= 
t (2 s+t) (k_1\cdot\ell)^2+2 (s+t)^2 k_1\cdot\ell\, k_2\cdot\ell
+(s+t)^2 k_2\cdot\ell^2-2 s^2 k_1\cdot\ell\, k_3\cdot\ell
\\&\hphantom{=}
- s^2 (k_3\cdot\ell)^2- s (s+t)^2 \ell^2\,,\\
\Numer_3 &= -2 t k_1\cdot\ell^2
-2 (s+t) k_1\cdot\ell\, k_2\cdot\ell
+2 (s-t) k_1\cdot\ell\, k_3\cdot\ell
+2 (s+t) k_2\cdot\ell\, k_3\cdot\ell
\\&\hphantom{=}
+2 s (k_3\cdot\ell)^2+s (s+t) \ell^2\,.
\end{aligned}
\end{equation}
As shown 
in Ref.~\cite{Gambuti:2023eqh} we can express all numerators
as combinations of Gram determinants.  These determinants 
are defined as follows,
\begin{equation}
	G\begin{pmatrix}
		q_1 &,\dots, & q_m\\
		p_1 &,\dots, & p_m\\
	\end{pmatrix}\equiv \det (2 q_i \cdot p_j)\, ;
\end{equation} 
where if the sequences are identical we list only one. 
From Ref.~\cite{Gambuti:2023eqh}, we need the following
determinants for the on-shell box,
\begin{equation}
G_1=G\begin{pmatrix}
   \loopm &k_1 & k_2 & k_3	
  \end{pmatrix},
G_2=G\begin{pmatrix}
   \loopm &k_1 & k_2 \\
   \loopm &k_3 & k_4	
\end{pmatrix},
G_3=G\begin{pmatrix}
   \loopm-k_1 &k_2 & k_3 \\
   \loopm &k_1 & k_4	
\end{pmatrix} \ .
\end{equation}
Defining the ideal~\cite{Gambuti:2023eqh}, 
\begin{equation}
J=\braket{G_1,G_2,G_3}\,,
\end{equation} 
with respect to the variables,
\begin{equation}
    \set{\ell\cdot k_1,\ell\cdot k_2,\ell\cdot k_4,\ell^2}\,,
\end{equation}
we find that,
\begin{equation}
    \Numer_i\! \mod J = 0\,,
\end{equation}
that is the numerators are expressible in terms of the
rank-two basis numerators from Ref.~\cite{Gambuti:2023eqh}.
(We find a similar equivalence of bases at rank 3.)

The attentive reader will notice that this immediately
poses a puzzle: we expect two finite integrands, matching
the two lattice points in the relative interior; but we find
three in the loop-momentum representation.  One clue to the
resolution of this puzzle may be found in the parametric
representation of the three integrals with $G_i$ as
numerators,
\begin{equation}
\begin{aligned}
\hat I[G_1] &= 2 s t (s+t) (D-3) \, 
   \Gamma\biggl(3-\frac D 2\biggr)
  \int \dd^3\feynp \, \mathcal F_{\onebox}^{-4+D/2}
	\mathcal U_{\onebox}^{2-D} 
    [t\feynp_ 2+s\feynp_1\feynp_ 3] \ , \\
\hat I[G_2] &=  s \, \Gamma\biggl(3-\frac D 2\biggr)
  \int \dd^3\feynp \, \mathcal F_{\onebox}^{-4+D/2} 
     \mathcal U_{\onebox}^{2-D} 
\\&\hspace*{25mm} \times
[\feynp_2 t (2(D-4) s + (3D-10) t)
+\feynp_1\feynp_3(D-2) s (s + 2 t)]
   \, , \\
\hat I[G_3] &= t\, \Gamma\biggl(3-\frac D 2\biggr)
  \int \dd^3\feynp \, \mathcal F_{\onebox}^{-4+D/2}
     \mathcal U_{\onebox}^{2-D}
\\&\hspace*{25mm} \times
[\feynp_2 (D-2) t (2 s + t) 
+\feynp_1\feynp_3 s ((3D-10) s + 2(D-4) t)]
   \, .
\end{aligned}
\end{equation}
The integrands depend only on two Feynman-parameter
monomials, $\set{\feynp_ 1, \feynp_ 1 \feynp_ 3}$.
Accordingly, there is a linear combination of the $G_i$
(or the $\Numer_i$) whose 
Feynman-parameter representation vanishes identically:
\begin{equation}
\begin{aligned}
\Numer_{\onebox,0} &= 
s t (2 s+t)\, k_1\cdot\ell-2 t^2 (k_1\cdot\ell)^2
+s t (s+t)\, k_2\cdot\ell-4 t (s+t)\, k_1\cdot\ell\, k_2\cdot\ell
\\&\hphantom{=} 
-2 (s+t)^2 (k_2\cdot\ell)^2+s^2 t\, k_3\cdot\ell
+4 s t\, k_1\cdot\ell\, k_3\cdot\ell
-4 s (s+t)\, k_2\cdot\ell\, k_3\cdot\ell
\\&\hphantom{=} 
-2 s^2 (k_3\cdot\ell)^2
+s t (s+t) \ell^2\,.
\end{aligned}
\label{OneLoopTotalDerivative}
\end{equation}
This combination is nontrivial in the loop
momentum (and $D$-independent).
It is locally finite, as it is a combination of locally
finite integrands.  To understand it better, we may note that 
the corresponding integrand is a total derivative,
\begin{equation}
  \label{total-derivative}
  \frac{\Numer_{\onebox,0}(\ell,\kset)}{\Den_1 \cdots \Den_4}
  = \frac{\partial}{\partial \ell^{\mu}} \, \frac{v^{\mu}}{\Den_1 \cdots \Den_4}\,,
  \qquad
  v^{\mu} = c_{2,1} v_{2,1}^{\mu} + c_{2,2} v_{2,2}^{\mu} + c_3 v_3^{\mu}\,,
\end{equation}
where \(v_{2,i}\) and \(v_3\) are IBP generating vectors 
(solutions of the corresponding syzygy 
equations~\cite{Gluza:2010ws,Georgoudis:2016wff,Wu:2023upw}) 
of degree two and three respectively, \(c_{2,i}\) are 
polynomials linear in~\(\ell\), and \(c_3\) is a constant 
coefficient.  The vectors and coefficients have
compact expressions in terms of the edge momenta,
\begin{equation}
  q_1 = \ell,
  \qquad
  q_2 = \ell - k_1,
  \qquad
  q_3 = \ell - k_1 - k_2,
  \qquad
  q_4 = \ell - k_1 - k_2 - k_3,
\end{equation}
manifesting the symmetries of the underlying graph (though 
at the cost of obscuring the degree in~\(\ell\)) as follows:
\begin{equation}
  \begin{aligned}
    v_{2,1}^\mu &= q_3^2 \, q_1^\mu - q_1^2 \, q_3^\mu,
    &
    c_{2,1} &= \frac{t^2}{2(D-3)} \left( q_1^2 - q_3^2 \right), 
    \\
    v_{2,2}^\mu &= q_4^2 \, q_2^\mu - q_2^2 q_4^\mu,
    &
    c_{2,2} &= \frac{s^2}{2(D-3)} \left( q_2^2 - q_4^2 \right), 
    \\
    v_3^\mu &= \left[ (q_2 \cdot q_4) q_3^2 - q_2^2 q_4^2 
        \right] q_1^\mu + \text{cyclic},
    &
    c_3 &= - \frac{st}{D-3}\,,
  \end{aligned}
\end{equation}
where ``cyclic'' stands for the three cyclic permutations of 
edge momentum indices.

\Eqn{total-derivative} shows that a nontrivial total derivative in momentum space may yield an identically zero integrand in Feynman-parameter space.  It also shows that we must
consider IBP identities in order to fully characterize
independent numerators yielding finite integrals.
The appearance of a total derivative
is an example of the subtleties that arise
in comparing numerators obtained using the Landau
analysis of Ref.~\cite{Gambuti:2023eqh} with the polytope
analysis we are developing in the present article.
A Feynman-parameter representation does not always distinguish 
total derivatives from locally finite integrands.
In the massless box case, such total derivative happens to be 
locally finite, therefore the sets of finite numerators agree 
in both approaches without any further analysis.
In the following sections, we will see that this is not always 
the case. This may be connected to Collins's observation \cite{Collins:2020euz} that a solution of Landau equations does not necessarily lead to a singularity in the parametric representation, but we have not investigated this question.

This discussion shows that solving the linear system of equations
\eqref{InteriorEquations} may in general lead to parametric numerators that vanish identically.
Where the set of relative-interior points is empty, 
this occurs trivially.
It can also occur in cases where this set of points is nonempty.

\subsection{Massless Double Box}
\begin{figure}[htb]
	\centering
	\begin{tikzpicture}
    \begin{scope}[massless]
		\draw(-1,-1)--+(-135:0.8) node[left] {$k_1$};
		\draw(3,-1)--+(-45:0.8) node[right] {$k_4$};
		\draw(3,1)--+(45:0.8) node[right] {$k_3$};
		\draw(-1,1)--+(135:0.8) node[left] {$k_2$};
		\draw(-1,-1)-- node[below] {$1$} (1,-1) -- node[below] {$6$} (3,-1) -- node[right] {$7$} (3,1) -- node[above] {$5$} (1,1) -- node[above] {$3$} (-1,1) -- node[left] {$2$} cycle;
		\draw(1,-1)-- node[right] {$4$} (1,1);
    \end{scope}
	\end{tikzpicture} 
	\caption{Double box with all massless internal and external lines}
	\label{double-box}
\end{figure}
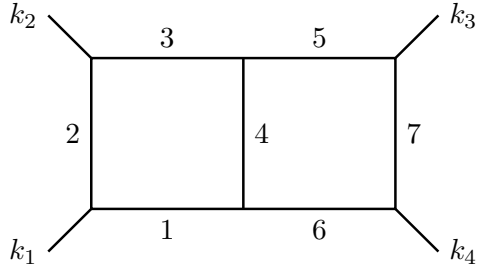

We consider next the massless double box shown in 
Fig.~\ref{double-box}.  This example allows for higher-rank 
numerators. 
The denominators associated with the graph are, 
\begin{equation}
\begin{aligned}
\Den_1&=\loopm_1^2,\quad
\Den_2=(\loopm_1-k_1)^2,\quad
\Den_3=(\loopm_1-K_{12})^2,\quad
\Den_4=(\loopm_1-\loopm_2)^2,
 \\ 
\Den_5&=(\loopm_2-K_{12})^2,\quad 
\Den_6=\loopm_2^2,\quad 
\Den_7=(\loopm_2-K_{123})^2\,,
\end{aligned}
\end{equation}
where $K_{ijk \dots}=k_i+k_j+k_k+\dots$.  (Note that
the order, and hence labels, of propagators is
different from Ref.~\cite{Gambuti:2023eqh}.)
\def\doublebox{{\square\mskip-2.5mu\square}}
\def\threeladder{{\square\mskip-2.5mu\square\mskip-2.5mu\square}}
The Symanzik polynomials are then, 
\begin{equation}
\begin{aligned}
\cU_{\doublebox}&=
\feynp_{123} \feynp_{4567}+\feynp_4 \feynp_{567}, 
\\ \cF_{\doublebox}&=-t \feynp_2 \feynp_4 \feynp_7
- s \left(\feynp_6 \left(\feynp_5 \feynp_{234}
+\feynp_3 \feynp_4\right)
+\feynp_1 \left(\feynp_3 \feynp_{4567}
+\feynp_5 \feynp_{46}\right)\right)\,.
\end{aligned}    
\end{equation}
We use the shorthand notation
$\feynp_{ijk\cdots}=\feynp_i+\feynp_j+\feynp_k+\cdots$. 
At rank zero, we have $n_\cU = -1$ when $D=4$, so we treat $\cU_{\doublebox}$ as the numerator.
The Newton polytope of the denominator, given by $\cF_{\doublebox}^3$, has no interior points, so the scalar integral diverges.
At rank one, the denominator is the same; all exponents of the numerator are therefore exterior.
Setting there coefficients to zero we find no solutions for the momentum-space ansatz, so we move to rank two. 
The parametric representation of 
the  Feynman integral with a numerator up to rank two reads 
\begin{align}
\hat I_{\doublebox}[\Numer(\pmb{\loopm}, \kset)]=
  \Gamma\left[6-D \right] 
  \sum_{\evec\in C_{\doublebox,2}} a_{\evec}  \int \dd^6 \feynp
  \cF_{\doublebox}^{-7+D}	\cU_{\doublebox}^{5-\frac 3 2 D} 
	\feynpset^\evec  \, .
\end{align}
For rank $r\ge 2$ we have, 
\begin{equation}
    P_{\doublebox,r} =  3 \,\Newton(\cF_\doublebox)
         +(r-1)\,\Newton(\cU_{\doublebox})\, .
\end{equation}
The $H$-representation~\eqref{HRepresentation} of the polytope
is given by the $A$ matrix, 
\setcounter{MaxMatrixCols}{20}
\begin{equation}
A_\doublebox^T= \left(	
\begin{matrix}
1 & 1 & 1 & 0 & 1 & 0 & 0 & 0 & 0 & 0 & 1 & -1 & 0 & 0 
& -1 & -1 & 0 & -1 \\
1 & 1 & 1 & 1 & 1 & 0 & 0 & 0 & 0 & 1 & 0 & -1 & -1 & 0 
& -1 & -1 & -1 & -1 \\
1 & 1 & 1 & 1 & 0 & 0 & 0 & 0 & 1 & 0 & 0 & 0 & -1 & 0 & 
-1 & 0 & -1 & -1 \\
1 & 1 & 1 & 0 & 0 & 0 & 0 & 1 & 0 & 0 & 0 & 0 & 0 & -1 & 0 & -1 & -1 & -1 \\
0 & 1 & 0 & 1 & 0 & 0 & 1 & 0 & 0 & 0 & 0 & 0 & 0 & 0 & 0 & 0 & -1 & -1 \\
1 & 0 & 0 & 0 & 1 & 1 & 0 & 0 & 0 & 0 & 0 & 0 & 0 & 0 & 0 & -1 & 0 & -1\\
\end{matrix}
\right)
\, ,
 \end{equation}
 and the $b$ vector,
\begin{equation}
 	-b_{\doublebox,r}^T= \begin{matrix}\big(
 \!-r\!-\!5\!&\!-r\!-\!5\!&\!-r\!-\!2\!&\!-3\!&\!-3\!&\!0\!
 &\!0\!&\!0\!&\!0\!&
 \\\hfill\quad\!0\!&\!0\!&r\!+\!2&r\!+\!2&r\!+\!2&
 r\!+\!5\!&\!2 (r\!+\!2)\!&\!2 (r\!+\!2)\!&\!2 (r\!+\!4)\!-\!1
 \big)
 \end{matrix}
 \, .
\end{equation}
The number of columns in these transposes corresponds to
the number of facets in the polytope. 
We start with seven Feynman parameters,
and set the last one to 1, so the polytope is embedded in
a six-dimensional space.
At rank 2, the Feynman polytope $P_{\doublebox,2}$
contains 1229 lattice points.
Of these, 117 can arise from Feynman integrands, 
so that $|C_{\doublebox,2}|=117$.  
Using algorithms N and G, we find $12$ points in the relative interior,
\begin{equation}
    \begin{aligned}
    R_{\doublebox,2}= &\big\{
(1,0,1,1,0,0),(0,1,0,2,0,0),(1,0,1,2,0,0),(1,0,1,1,1,0),\\
&\hphantom{\big\{}
(1,0,0,2,1,0),(1,0,1,1,0,1),(0,0,1,2,0,1),(1,0,1,0,1,1),
\\
&\hphantom{\big\{}
(1,0,0,1,1,1),(0,1,0,1,1,1),(0,0,1,1,1,1),(0,0,0,2,1,1)
\big\}\,.
    \end{aligned}
    \label{interior2box}
\end{equation}
We must set to zero the coefficients of the monomials in 
$C_{\doublebox,2}$ that are not in $R_{\doublebox,2}$.
Here this means solving a system of 105 equations. 
As we increase the rank, the number of equations
increases, and we need an efficient way of solving the systems. 
Finite-field methods are a convenient tool; we use 
the \textsf{FiniteFlow} package \cite{Peraro:2019svx}.
(We use the built-in command \textsf{FFDenseSolve}.)
This yields two finite numerators at rank two. 
This number coincides with the one obtained in 
Ref.~\cite{Gambuti:2023eqh}, with Gram generators,
\begin{equation}
	G_{\doublebox,1}=G\begin{pmatrix}
		\loopm_1 &k_1 & k_2\\
		\loopm_2& k_3 & k_4
	\end{pmatrix} , \quad
	G_{\doublebox,2}=G\begin{pmatrix}
		\loopm_1 &k_1 & k_2\\
		k_1& k_2 & k_4
	\end{pmatrix} 
	G\begin{pmatrix}
		\loopm_2 &k_3 & k_4\\
		k_1& k_2 & k_4
	\end{pmatrix}\,.
\end{equation}
\begin{table}[tb]
\centering
\setlength{\tabcolsep}{1em}
\begin{tabular}{llllll}
  \diagbox[width=5em]{$L$}{rank}
  & $1$  &  $2$ & $3$ & $4$ &$5$  \\
  \hline
  1& 0	& 2 & 8 & 19&  36 \\
  2& 0 & 12 & 135& 644&2095
  \\
  3 & 149 & 149 & 3684& 31863 & 167710\\
\end{tabular}
\caption{The number of relative-interior points 
 at different ranks in the Feynman polytope  
 $P_{F,r}$ 
 for the $L$-loop ladder.  These numbers were computed using
 both algorithm G based on \Eqn{gen-function} and algorithm N.}
	\label{results-monomials}
\end{table} 
Upon conversion to the Feynman-parameter representation,
as expected these polynomials are expressed solely in terms of
monomials with exponents in the relative interior
$R_{\doublebox,2}$.
We indeed find that the solutions
to the polytope system are expressible as linear
combinations of these Gram generators.  At rank two,
the `locally finite' (or `Landau finite') approach 
of Ref.~\cite{Gambuti:2023eqh} and the 
`polytopically finite' approach we are developing here
thus give the same set of finite numerators.
The number of relative interior points at each rank is
summarized in Table \ref{results-monomials}.
At rank four we find that the numerator,
\begin{align}
	\Numer_{r=4}=(\loopm_1-k_1)^2(\loopm_2+k_4)^2 \, ,
\end{align}
depends on all 644 Feynman-parameter monomials, so that 
loop-momentum polynomials
cover the full set of relative interior points 
in the parametric representation.

Following our algorithm to construct numerators and 
both Algorithms N and G to compute interior points  
in Sect.~\ref{algorithm}, we find agreement
between the locally finite and polytopically finite
approaches through rank four.  We start with the generators
given in Ref.~\cite{Gambuti:2023eqh}, and check that all numerators in the two approaches
span the same linear space. 

\newcommand{\Nlf}{\mathcal{V}_{\mathrm{LF}}}
\newcommand{\Npf}{\mathcal{V}_{\mathrm{PF}}}
\newcommand{\Ntd}{\mathcal{V}_{\mathrm{TD}}}
\newcommand{\Ilf}{I_{\mathrm{LF}}}
\newcommand{\Ipf}{I_{\mathrm{PF}}}
\newcommand{\IBP}{\mathrm{IBP}}

At rank five, we find 247 independent loop-momentum 
numerators using the Landau analysis versus 313 obtained with 
the polytopic analysis.
Let us call the corresponding linear spaces $\Nlf$ and $\Npf$.
To understand the origin of this mismatch, we compare
the linear spaces directly.%
\footnote{We use \textsf{FiniteFlow}~\cite{Peraro:2019svx} 
to speed up linear algebra computations with coefficients 
depending on parameters.}
This gives \(\Nlf \subset \Npf\), in agreement with the 
expectation that absolutely convergent momentum-space integrals 
yield absolutely convergent parameter-space integrals (see 
Appendix~{app:abs-conv}).

We find that all 313 independent numerators obtained
in the polytopic analysis are locally IR finite.
What is the source of the excess in this analysis?
To understand it, there is another issue to resolve,
that of UV convergence.

The Landau analysis as implemented in 
Ref.~\cite{Gambuti:2023eqh} imposed a \textit{strong\/} UV
finiteness constraint, requiring overall and per-loop
UV finiteness by power counting.  However, there are
numerators which violate these conditions but where the
coefficient of the would-be UV divergence vanishes
(for example, by being proportional to a vanishing
invariant).  We can call this \textit{simply weak\/} UV 
convergence\footnote{The coefficient of the would-be
divergence can also be proportional to the dimensional
regulator $\eps$, in which case we would speak of
\textit{evanescently\/} weak UV convergence.}.
Consider the following numerator,
\begin{equation}
    \Numer =
    \left( \loopm_1 \cdot k_1 \right)
    \left( \loopm_1 \cdot k_2 \right)^2
    G_{\doublebox, 2} \,.
\label{WeakConvergenceExample}
\end{equation}
It fails the power-counting criterion for the $\loopm_1$
loop.  However, in dimensional regularization (or using
another UV regulator consistent with Lorentz invariance),
we find that the leading UV behavior of
the integration over $\loopm_1$ is, 
\begin{equation}
  \int \dd^4 \loopm_1
  \frac{\loopm_1^\mu \loopm_1^\nu \loopm_1^\rho \loopm_1^\sigma}
  {\left( \loopm_1^2 \right)^4}
  \propto \eta^{\mu \nu} \eta^{\rho \sigma} 
  + \eta^{\mu \rho} \eta^{\nu \sigma} 
  + \eta^{\mu \sigma} \eta^{\nu \rho}
\end{equation}
contracted with external momenta and \(\loopm_2\).
This contraction may vanish, as indeed happens for the
example in \Eqn{WeakConvergenceExample}.
In this case the regulated integral is finite.

Rank five is the lowest rank at which this distinction matters
for locally IR finite numerators in the double box.  
We indeed find that
imposing strong UV finiteness on the polytopically finite
numerators we obtain the same linear space as in
the Landau analysis.

This is not the end of comparisons, however.  If we replace
the strong UV convergence imposed in Ref.~\cite{Gambuti:2023eqh}
on the Landau analysis
with simply weak UV convergence, we find a linear space
spanned by 307 independent numerators, six fewer than
in the polytopic analysis.  An example of a numerator
which appears in the latter analysis but is excluded
even by the modified Landau analysis is the following,
\begin{equation}
    \Numer =
    \left( \loopm_1 \cdot k_1 \right)
    \left( \loopm_1 \cdot k_3 \right)^2
    G_{\doublebox, 2}\,.
\end{equation}
As we have seen in the previous section, the additional numerators in~\(\Npf\) may be related to the ones in~\(\Nlf\) by total derivatives.  This is the case for the six
additional numerators: the corresponding
integrands are all linear combinations
of weakly UV finite integrands arising from the modified
Landau analysis and of total derivatives.  We can show
this by reducing both sets of integrands by standard
IBP reductions.  More precisely, we compare the linear
spaces with $D$-independent coefficients after reduction
to master integrals,
and find they are the same.  
In the case studied here, it suffices to use total
derivatives of the same numerator rank as the numerators
under study.  This is not the case in general.
(We remind the reader that the
independent numerators found in Ref.~\cite{Gambuti:2023eqh} and
here do not all yield independent master integrals, as
IBP relations have not been taken into account.)

\subsection{Nonplanar Double Box}
\begin{figure}[ht]
	\centering
	\begin{tikzpicture}
    \begin{scope}[massless]
		\draw(-1,-1)--+(-135:0.8) node[left] {$k_1$};
		\draw(-1,1)--+(135:0.8) node[left] {$k_2$};
		\draw(-1,1)-- node[above] {$3$} (1,1) -- node[above] {$5$} (3,1);
		\draw(-1,-1)-- node[below] {$1$} (1,-1) -- node[below] {$6$} (3,-1);
		\draw(-1,-1)-- node[left] {$2$} (-1,1);
		\draw(1,-1)--(1.95,-0.05);
		\draw(2.05, 0.05)-- node[pos=0.6,anchor=north west] {$7$} (3,1);
		\draw(3,-1)-- node[pos=0.8,anchor=north east] {$4$} (1,1);
    \end{scope}
    \pgfsetblendmode{multiply}
    \begin{scope}[transparency group]
		\draw[massive] ($(3,-1)-(-22.5:0.05)$)-- +(-22.5:0.85) node[right] {$k_4$};
		\draw[massive] ($(3,1)-(22.5:0.05)$)-- +(22.5:0.85) node[right] {$k_3$};
    \end{scope}
	\end{tikzpicture} 
	\caption{Nonplanar double box with dashed legs off-shell}
	\label{nonplanar-double-box}
\end{figure}
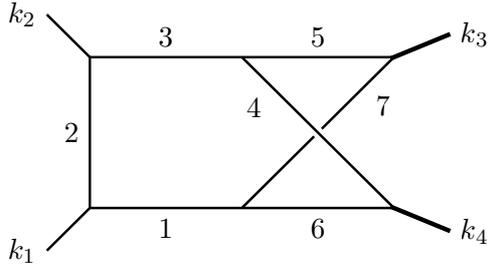

\def\npdbox{\times}

In previous sections, we considered planar integrals.
Planar integrals can be considered in Euclidean
kinematics, where the Symanzik polynomials are positive
away from the boundaries of $\mathbb R ^E_{>0}$,
and hence
completely non-vanishing in the sense of
Sect.~\ref{convergence}.  The integrals in this
region then satisfy the preconditions of the BFP
theorem.
In this section, we consider the nonplanar double box,
shown in Fig.~\ref{nonplanar-double-box}.
With a judicious choice of external masses, we can still
retain the positivity property and satisfy the
preconditions of the BFP theorem.  We will consider
both this choice as well as the massless limit.  The
denominators are,
\begin{equation}
\begin{aligned}
  \Den_1=& (\loopm_1+K_{12})^2,\quad 
  \Den_2= (\loopm_1+k_2)^2, \quad
  \Den_3= \loopm_1^2,\quad 
  \Den_4=(\loopm_1-\loopm_2)^2,\quad 
  \\ & 
  \Den_5=\loopm_2^2,\quad 
  \Den_6=(\loopm_1-\loopm_2-k_4)^2, \quad
  \Den_7= (\loopm_2-k_3)^2 \,.
\end{aligned}
\end{equation}
We set $k_3^2=k_4^2=m^2$ so that $m^2=(s+t+u)/2$. The first Symanzik polynomial reads,
\begin{equation}
\cU_{\npdbox}=
  \feynp_1 \feynp_{4567}+\feynp_2 \feynp_{4567}
  +\feynp_7 \feynp_{634}+\feynp_3 \feynp_{645}
  +\feynp_4 \feynp_5+\feynp_6 \feynp_5 \, ,
\end{equation}
and the second Symanzik polynomial is,
\begin{equation}
\begin{aligned}
 \cF_{\npdbox}= &
  -\frac  {s} {2} [
  \feynp_6 \left(\feynp_4 \feynp_{23}
  +\feynp_5 \feynp_{34}\right)
  +\feynp_7 \left(\feynp_5 \feynp_{24}
  +\feynp_6 \feynp_{45}
  +\feynp_3 \big(\feynp_{456}
  + \feynp_6\right)\big)
  \\&\hspace{10mm}
  +\feynp_1 \big(\feynp_5 \feynp_{67}
  +\feynp_4 \left(\feynp_{567}+ \feynp_5\right)
  +2 \feynp_3 \feynp_{4567}\big)]
   \\
	&-\frac  {t} {2} [\feynp_6 \left(\feynp_4 \feynp_{312}+\feynp_5 \feynp_{413}\right)+\feynp_7 \left(\feynp_{45} \feynp_{613}+\feynp_4 \feynp_5+\feynp_2 \left(2 \feynp_4+\feynp_5\right)\right)]\nonumber
	\\
	&-\frac{u}{2}[\feynp_7 \left(\feynp_5 \feynp_{24}+\feynp_{45} \feynp_{613}\right)+\feynp_6 \left(\feynp_4 \feynp_{312}+\feynp_5 \left(\feynp_{413}+2 \feynp_2\right)\right)] \, .
\end{aligned} 
\label{nonplanardoublebox}
\end{equation}
In the massless limit, we have $u=-s-t$,
so that coefficients of the monomials on the third line of \Eqn{nonplanardoublebox} become positive and 
accordingly $\cF_{\npdbox}$ is no longer of definite sign.
\subsubsection{Massive case}
Let us consider the massive case first.
With $D=4$ and $r=0$, we have $n_\cU = -1$, so the numerator is simply $\cU_{\npdbox}$.
Most of its monomials (14 out of 16) are exterior w.r.t.\ the Newton polytope of the denominator, so the scalar integral diverges.
At rank one, we have $n_\cU=0$, $n_\cF=3$; hence, 
\begin{equation}
P_{\npdbox}= 3\, \Newton(\cF)\,,
\end{equation}
and using algorithms N,G, we find the relative interior points,
\begin{equation}
R_{\npdbox,1} = \set{(0,0,0,1,0,0),(0,0,0,0,1,0),
(0,0,0,1,0,1),(0,0,0,0,1,1)} \,. 
\end{equation}
At this rank, there are $|C_{\npdbox,1}|=16$ monomials
that can arise in Feynman integrals, corresponding to
the points,
\begin{equation}
\begin{aligned}
C_{\npdbox,1} &=\{(1, 0, 0, 1, 0, 0), (1, 0, 0, 0, 1, 0), 
(1, 0, 0, 0, 0, 1), (1, 0, 0, 0, 0, 0), 
\\&\hphantom{=\ \{}
(0, 1, 0, 1, 0, 0), (0, 1, 0, 0, 1, 0), (0, 1, 0, 0, 0, 1), 
(0, 1, 0, 0, 0, 0), 
\\&\hphantom{=\ \{}
(0, 0, 1, 1, 0, 0), (0, 0, 1, 0, 1, 0), (0, 0, 1, 0, 0, 1),
(0, 0, 1, 0, 0, 0), 
\\&\hphantom{=\ \{}
(0, 0, 0, 1, 1, 0), (0, 0, 0, 1, 0, 0), (0, 0, 0, 0, 1, 1), 
(0, 0, 0, 0, 0, 1)\} \,.
\end{aligned}
\end{equation}
Following our FNA algorithm we find a lone finite numerator,
\begin{equation}
	\Numer = \frac{2}{s}[
	s \left(-\loopm_1\cdot k_4+ \loopm_1\cdot k_3\right)+t \left(\loopm_1\cdot k_2-\loopm_1\cdot k_1\right)+u \left(\loopm_1\cdot k_1-\loopm_1\cdot k_2\right)] \,.
\end{equation}
At ranks two, three, and four, we find 10, 49, and 174 
independent numerators respectively.
This agrees with the Landau analysis of Ref.~\cite{Gambuti:2023eqh} for this integral.
At rank five we find $504$ independent numerators 
in the polytopic analysis versus $469$ numerators obtained using the Landau analysis. However, relaxing the strong UV convergence constraint we find full agreement between the two approaches.

\subsubsection{Massless case}
If we set $m=0$ in \Eqn{nonplanardoublebox}, there are
both negative and positive terms in $\cF$. This means that the polynomial vanishes for some positive $\feynp$s, and the integral may diverge somewhere inside the integration domain.
The BFP theorem no longer applies in this case, as it only probes divergences on the boundary, that is, when a subset of $\feynp$s goes to $0$ and/or $\infty$.
However, explicit solution of the Landau equations for the nonplanar double box shows that there are in fact no divergences inside the integration domain~\cite{Gardi:2022khw,Gambuti:2023eqh}.

This analysis tempts us to apply our algorithm even
in the absence of strict satisfaction of the BFP preconditions.
At ranks zero through two, there are no finite numerators.
At ranks three, four, and five, we find respectively
$9$, $65$, and 263 finite numerators. This agrees with 
spaces found in the Landau analysis. 

The analyses of Refs.~\cite{Gardi:2022khw,Gardi:2024axt} suggest that for two-loop $2 \to 2$ scattering Landau singularities live on the boundary of the integration domain. The GHJM
analysis~\cite{Gardi:2024axt} showed this no longer holds at 
three loops, where there are singularities associated with the 
interior as in the case of
Landshoff scattering. 

\subsection{Beetle}
\newcommand{\beetlegraph}{
\begin{tikzpicture}[thick,scale=0.1]
	\draw(-1,-0)--(0,-1);
	\draw[thick](-1,-1)--(-1.0,1.0)--(1.0,1.0)--(1.0,-1.0)--cycle;
\end{tikzpicture}
}
\def\beetle{{\,\raisebox{0.8pt}{\scaleto{\diagdown}{3.5pt}}%
\mskip-7.8mu\square}}
We turn next to the `beetle' graph, depicted
in Fig.~\ref{beetle-graph}. The denominators are, 
\begin{figure}
	\centering
\begin{tikzpicture}
    \begin{scope}[massless]
	\draw(-1,-1)--+(-135:0.8) node[left] {$k_1$};
	\draw(-1,-0)-- node[anchor=south west] {$7$} (0,-1);
	\draw(1,-1)--+(-45:0.8) node[right] {$k_4$};
	\draw(1,1)--+(45:0.8) node[right] {$k_3$};
	\draw(-1,1)--+(135:0.8) node[left] {$k_2$};
	\draw(-1,-1)-- node[left] {$2$} (-1,0)-- node[left] {$3$} (-1.0,1.0)-- node[above] {$4$} (1.0,1.0)-- node[right] {$5$} (1.0,-1.0)-- node[below] {$6$} (0,-1) -- node[below] {$1$} cycle;
    \end{scope}
\end{tikzpicture}
\caption{The beetle graph}
\label{beetle-graph}
\end{figure}
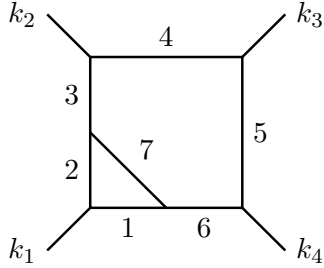
\begin{equation}
\begin{aligned}
	\Den_1&=\loopm_1^2,\quad
	\Den_2=(\loopm_1-k_{1})^2,\quad
	\Den_3=(\loopm_2-k_1)^2, \quad
	 \Den_4=(\loopm_1-K_{12})^2,
	\\ \Den_5&=(\loopm_1-K_{123})^2, \quad 
	\Den_6=\loopm_2^2, \quad
	\Den_7=(\loopm_1-\ell_2)^2\quad
\end{aligned}
\end{equation}
 The Symanzik polynomials read,
\begin{equation}
\begin{aligned}
\cU_{\beetle}&=\feynp_7 \feynp_{123456}
+\feynp_{36} \feynp_ {1245}, \\
\cF_{\beetle}&=-
\alpha _4 s \left(\feynp_1 \feynp_ {367} 
+\feynp_6 \feynp_7\right)
-t\feynp_5 \left(\feynp_3 \feynp_{7}
    +\feynp_2 \feynp_ {367}\right) \, ,
	\end{aligned} 
\end{equation}
where the kinematic invariants $s$ and $t$ are defined as in the massless box.
At ranks zero and one,
we find an empty relative interior, and there
are no finite numerators.  At rank two, we similarly have an empty interior but the solution of the system of equations leads to 
one finite numerator. Indeed, the system  of equations instruct us to take the coefficients of the monomials which are not in the interior to zero so they  must correspond to a vanishing numerator when there are no interior points. Indeed, we find, 
\begin{equation}
\Numer =
\left( \loopm_1 \cdot k_1 \right)
\left(
s \left( \loopm_2 \cdot k_3 \right)
- u \left( \loopm_2 \cdot k_2 \right)
\right)
- \left( \loopm_2 \cdot k_1 \right)
\left(
s \left( \loopm_1 \cdot k_3 \right)
- u \left( \loopm_1 \cdot k_2 \right)
\right) \, . 
\label{BeetleRank2}
\end{equation}
The Landau analysis, in contrast, yields no numerators
at this rank.  As it is linear in $\loopm_2$, the
expression~\eqref{BeetleRank2} is
UV convergent by power counting.  The explanation
for its appearance is simple: like the expression
in \Eqn{OneLoopTotalDerivative}, it vanishes
identically after conversion to Feynman parameters, and
is accordingly a total derivative. 
These solutions indeed can be put together in a vector space of 
numerators whose parametric representation vanishes identically.
We find the first notrivial polytopic numerators at rank three,
where we find 35 interior points and 12 numerators.

At ranks three, four, and five, we find 12, 39, and 90
independent numerators
from the polytopic analysis
after imposing the strong UV convergence constraint.
In contrast, we find 6, 29, and 76
independent numerators
from the Landau analysis of Ref.~\cite{Gambuti:2023eqh}.
Here, the comparison after eliminating total derivatives
is more subtle, because we need total derivatives of
higher numerator rank than that of the polytopic
numerators themselves at rank three.  We find the following
sequence of linear spaces of $D$-independent coefficients
after elimination of total derivatives,
\begin{equation}
\Nlf^{(3)} \subset \Npf^{(3)} \subset \Nlf^{(4)}
= \Npf^{(4)} \, , 
\end{equation}
where the superscripts denote the rank.  At rank
five, the two spaces are again identical.
\begin{table}[tb]
\centering
\setlength{\tabcolsep}{1em}
\begin{tabular}{@{}lrrr@{}}
  rank & 3 & 4 & 5 \\
  \midrule
  $\Nlf$ & 6 & 29 & 76 \\
  $\Npf$ & 12 & 39 & 90  \\ 
  $\Nlf / \Ntd$ & 4 & 14&  20 \\
  $\Npf / \Ntd$ & 5 & 14& 20  \\ 
\end{tabular}
\caption{The dimensions of spaces of independent
numerators before and after removal of total derivatives 
in the beetle graph.}
	\label{SpaceDimensions}
\end{table} 
The dimensions of the spaces after total-derivative
removal are given in Table~\ref{SpaceDimensions}.
With this additional subtlety in hand,
the polytopic analysis again produces the same numerators
as the Landau analysis, up to total derivatives.

\subsection{Three-Loop Ladder}
\begin{figure}[ht]
	\centering
	\begin{tikzpicture}
    \begin{scope}[massless]
		\draw(-1,-1)--+(-135:0.8) node[left] {$k_1$};
		\draw(5,-1)--+(-45:0.8) node[right] {$k_4$};
		\draw(5,1)--+(45:0.8) node[right] {$k_3$};
		\draw(-1,1)--+(135:0.8) node[left] {$k_2$};
		\draw(-1,1)-- node[above] {$7$} (1,1) -- node[above] {$6$} (3,1) -- node[above] {$5$} (5,1);
		\draw(-1,-1)-- node[below] {$1$} (1,-1) -- node[below] {$2$} (3,-1) -- node[below] {$3$} (5,-1);
		\draw(-1,-1)-- node[left] {$8$} (-1,1);
		\draw(1,-1)-- node[right] {$9$} (1,1);
		\draw(3,-1)-- node[right] {$10$} (3,1);
		\draw(5,-1)-- node[right] {$4$} (5,1);
    \end{scope}
	\end{tikzpicture} 
	\caption{Three-loop ladder with all massless internal and external lines.}
	\label{triple-box}
\end{figure}
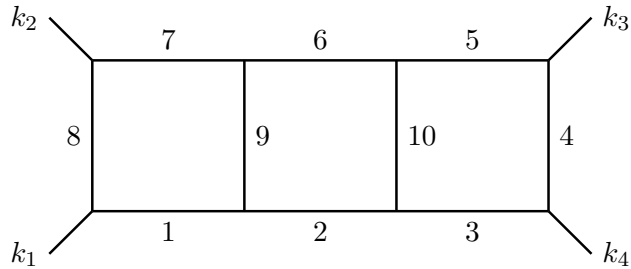
As our last example we consider the massless three-loop ladder 
integral, shown in Fig.~\ref{triple-box}. The denominators
for this graph are,
\begin{equation}
\begin{aligned}
\Den_1&= \loopm_1^2, \quad \Den_2=\loopm_2^2, \quad 
 \Den_3=\loopm_3^2, \quad  \Den_4=(\loopm_3 + k_4)^2, \quad 
 \Den_5=(\loopm_3-K_{12})^2, \\ 
\Den_6&=(\loopm_2 - K_{12})^2,\quad 
\Den_7=(\loopm_1 - K_{12})^2, \quad 
\Den_8=(\loopm_1 - k_1)^2,  \quad    
\Den_9=(\loopm_1-\loopm_2)^2,\quad  \\
\Den_{10}&= (\loopm_2-\loopm_3)^2
	\,,   
\end{aligned}
\end{equation}
where $k_i^2=0$. The Symanzik polynomials are, 
\begin{equation}
\begin{aligned}
\cU_{\threeladder} =&\,
\feynp_{10} \left(\feynp_9 \feynp_{12345678}
+\feynp_{178} \feynp_{23456}\right)
+\feynp_{345} \left(\feynp_9 \feynp_{12678}
+\feynp_{178} \feynp_{26}\right),
\\
\cF_{\threeladder}=& -t \feynp_4 \feynp_8 \feynp_9 \feynp_{10}  
-s [(\feynp_1 (\feynp_2 \left(\feynp_{3,10} \feynp_{567}
+\feynp_{45} \feynp_{67}\right)
+\feynp_{45} \left(\feynp_6 \feynp_{79}
+\feynp_7 \feynp_9\right)
\\&
+\feynp_{10} \left(\feynp_7 \feynp_{456}
+\feynp_9 \feynp_{567}\right)
+\feynp_3 \left(\feynp_5 \feynp_{6910}
+\feynp_6 \feynp_{79,10}+\feynp_7 \feynp_{9,10}\right))
\\&
+\feynp_2 \left(\feynp_{3,10} \left(\feynp_{56} \feynp_{78}
+\feynp_9 \feynp_{567}\right)
+\feynp_6 \feynp_{45} \feynp_{789}
+\feynp_7 \feynp_9 \feynp_{45}\right)
\\&
+\feynp_3 \left(\left(\feynp_{10} \feynp_{56}
+\feynp_5 \feynp_6\right) \feynp_{789}
+\feynp_5 \feynp_9 \feynp_{78}
+\feynp_7 \feynp_9 \feynp_{10}\right)] \, .
\end{aligned}    
\end{equation}
At ranks zero and one, we have $n_\cU<0$ so we consider $\cU$ as 
part of the numerator; we do not find any finite numerators in 
these cases.
At rank two we obtain the 
polytope $P_{\threeladder,2} = 4\, \Newton(\cF_{\threeladder})$.
We find using Algorithms N,G, 149 points in the relative interior,
and two independent finite numerators.  This agrees
with the Landau analysis. 
The calculation of interior points at ranks three to five becomes 
expensive using algorithm N in comparison with Algorithm G but we 
find agreement between both approaches.
As usual in the parametric representation, the complexity of 
the problem grows with the number of edges of the graph.
It also increases when there are numerators, with
the increasing number of terms in the resulting polynomials.

We have also
computed the finite numerators at ranks three and four.
Here, the Landau analysis finds a subspace of that
found by the polytopic analysis, $\Nlf^{(r)}\subset \Npf^{(r)}$.
However, after removing the subspace of  total derivatives
we again find agreement between the Landau and polytopic
analyses.  
\begin{table}[tb]
\centering
\setlength{\tabcolsep}{1em}
\begin{tabular}{@{}lrrr@{}}
  rank & 2 & 3 & 4 \\
  \midrule
  $\Nlf$ & 2 & 26 & 184 \\
  \addlinespace
  $\Npf$ & 2 & 30 & 218 \\ 
  \addlinespace
  $\Nlf / \Ntd$ & 2 & 8 & 42 \\
  \addlinespace
  $\Npf / \Ntd$ & 2 & 8 & 42 \\ 
\end{tabular}
\caption{The dimensions of spaces of independent
numerators before and after removal of total derivatives 
in the three-loop ladder graph.}
	\label{SpaceDimensionsLadder}
\end{table} 

\section{Conclusions}
\label{the-end}

In this article, we have studied the problem of finding all 
numerators
which give rise to finite Feynman integrals.  We have made use of
Newton polytopes and the BFP theorem to this end.  Feynman integrals
whose exponents of
numerator monomials in the parametric representation all lie on
relative interior points of a certain polytope are 
necessarily finite
according to the theorem.  We conjectured that this is also a
necessary condition.  We have tested this conjecture by comparing
with the GKNT Landau approach~\cite{Gambuti:2023eqh}, finding
agreement with the numerators obtained in the latter approach.  
The comparison requires attention to the details of the ultraviolet
convergence constraint, and taking into account integration-by-parts
identities.  We have further motivated the conjecture 
in Appendix~\ref{app:several-mons}

The agreement is also reflected in the match between the Gr\"obner bases
of numerators, and correspondingly that all exponent vectors for
the numerators obtained in Ref.~\cite{Gambuti:2023eqh} lie in the
relative interior of the polytope associated to the Feynman integrals.
The comparison is straightforward for one- and two-loop examples,
and more subtle for the three-loop ladder.   Our analysis can be
performed in the standard representation or the LP representation,
with consistent results.

Planar integrals generally satisfy all conditions for application of
the BFP theorem.  This is not true for nonplanar integrals, which
may have second Symanzik polynomials with terms of different sign
(related to the absence of a pure Euclidean region).  In the case
of the nonplanar double box, this is not essential as there are
no singularities inside the region of parametric integration~\cite{Gardi:2024axt}, and the results of the BFP
theorem still hold.  For more complicated integrals, such as the
three-loop example studied in Ref.~\cite{Gardi:2024axt}, the
interior does contain singularities.  We expect that breaking
up the region as discussed there would allow our analysis to be
applied separately to each resulting integral, and that finiteness
would require finiteness of the integrations over each separate
region.
It may be worthwhile to investigate the connections to tropical
geometry.  It will be interesting to explore an extension of
polytopes from four-dimensional to $D$-dimensional integrals,
and to see whether that makes possible a classification of
divergent integrals as well.

\section*{Acknowledgments}

We thank Michael Borinsky, Einan Gardi, and Sebastian Mizera for discussions. 
This work is  supported by 
the European Research Council under grant ERC--AdG--885414. 
DAK and LDLC would like to thank FAPESP grant 2021/14335-0 for support during an August 2023 workshop at SAIFR in Sao Paulo.
DAK would like to thank the Galileo Galilei Institute,
where this work was completed, for its support.

\appendix

\section{Feynman Parametrization of Integrals 
with Numerators} 
\label{tensor-parameters}
Let us consider integrals with a single-monomial
numerator of rank $r$,
\begin{equation}
\hat I_{j_1 \dots j_r}= \int \frac{\dd^D \loopm_1}{i\pi^{D/2}} 
\cdots \frac{\dd^D \loopm_L}{i\pi^{D/2}} \quad 
\frac{\loopm_{j_1}\cdot v_1 \, \cdots
\, \loopm_{j_r}\cdot v_r}{\Den_1 \cdots \Den_E}\,,	
\end{equation}
where we have contracted $r$ loop momenta with generic vectors $v_1, \dots, v_r$ to make it scalar. 
Introducing Schwinger parameters we obtain the 
following expression,
\begin{equation}
 \begin{aligned}
 \hat I_{j_1 \dots j_r} =
 i^{-E} &\int \dd T \,T^{E-1} \int  \dd^E \alpha \,
\delta(1- \sum_{i=1}^E \alpha_i) e^{-i T \mathcal F/ \mathcal U }
	\int \frac{\dd^D \loopm_1}{i\pi^{D/2}} 
\cdots \frac{\dd^D \loopm_L}{i\pi^{D/2}} 
     e^{i T  \widetilde M_{ij}\loopm_i\cdot\loopm_j}
     \\&\times
 [\loopm_{j_1}\cdot v_1+(\widetilde M^{-1})_{j_1 j}Q_j\cdot v_1]
\cdots[\loopm_{j_r}\cdot v_r+(\widetilde M^{-1})_{j_r j} \widetilde Q_j\cdot v_r]\,.
 \end{aligned}
 \label{tensor-integral-pre-loop}
\end{equation}
We define the loop integral, 
\begin{equation}
Z[J] = \int \frac{\dd^D \loopm_1}{i \pi^{D/2}} \cdots 
\frac{\dd^D \loopm_L}{ i \pi^{D/2}} 
e^{i T  \widetilde M_{ij} \loopm_i\cdot \loopm_j
     +i J_i\cdot \loopm_i  }= 
 e^{-i (\widetilde M^{-1})_{lm} J_l\cdot J_m/(4 T) }Z[0],
\end{equation}
where 
\begin{equation}
Z[0]=\int \frac{\dd^D \loopm_1}{i \pi^{D/2}} \cdots 
\frac{\dd^D \loopm_L}{i \pi^{D/2}} 
\;e^{i T \ell_i^\mu  \widetilde M_{ij} \eta_{\mu\nu} \ell_j^\nu} = (i T)^{-DL/2}(\mathcal U)^{-D/2} \, . 
    \end{equation}
After expanding the product in the second line of \Eqn{tensor-integral-pre-loop}, 
the loop integrals we want to express have the form, 
\begin{equation}
v_{1,\mu_1}	
\cdots v_{m,\mu_m}	\braket{\loopm_{j_1}^{\mu_1}
		\cdots k_{j_m}^{\mu_m}} \equiv
\int \frac{\dd^D \loopm_1}{i \pi^{D/2}} \cdots 
\frac{\dd^D \loopm_L}{i \pi^{D/2}} 
\;e^{i T \widetilde M_{ij} \loopm_i\cdot\loopm_j}
\loopm_{j_1}\cdot v_1\cdots \loopm_{j_m}\cdot v_m\,,
\end{equation}
which we can compute using the generating functional by taking derivatives. 
For example, 
\begin{equation}
v_{1,\mu}	 v_{2,\nu}	\braket{\loopm_{j_1}^{\mu}
\loopm_{j_2}^{\nu}}=\frac{i}{2 T} (\widetilde M^{-1})_{j_1 j_2} 
 v_1\cdot v_2 \,Z[0]=
 \frac{i^{1-DL/2}}{2}	(\widetilde M^{\text{adj}})_{j_1 j_2}  
 (\mathcal U)^{-D/2-1} T^{-DL/2-1} v_1\cdot v_2\,.
 \label{example-wick-2}
\end{equation}
Here we have used 
$(\widetilde M^{-1})_{j_1j_2}=(\cU)^{-1} \widetilde M^{\text{adj}}_{j_1j_2}$, 
where "$\text{adj}$" stands for the adjugate matrix of $\widetilde M$ 
(the transpose of its cofactor).
Higher-rank integrals can be obtained from Wick's theorem, 
keeping in mind that contractions of an odd number of vectors 
vanish. The remaining integral over $T$ is trivial.
Using Wick contractions and solving the integral over $T$ leads to the parametric representation of the tensor integral over a single term. 
For instance, the rank-one case  is
given by, 
\begin{equation}
    \hat I_{j_1}=
 i^{-E} \int \dd T \,T^{E-1} \int  \dd^E \alpha \,
\delta(1- \sum_{i=1}^E \alpha_i) e^{-i T \mathcal F/ \mathcal U } 
(\widetilde M^{\text{adj}})_{\sigma_1 j}v_1\cdot Q_j Z[0],
\end{equation}
so that, 
\begin{align}
	\hat I_{j_1}=(-1)^E \Gamma[E-DL/2] \int  \dd^E \alpha \,
	\delta(1- \sum_{i=1}^E \alpha_i) \, (\widetilde M^{\text{adj}})_{\sigma_1 j}v_1\cdot Q_j\ 
	\mathcal U^{N-D/2(L+1)-1}
	\mathcal{F}^{DL/2-N}.
\end{align}
after integration over $T$.
Similarly, using \Eqn{example-wick-2} the rank-two integral is, 
\begin{equation}
\begin{aligned}
    	\hat I_{j_1 j_2}= &-      i^{-2 E} \frac 1 2 
     \Gamma[E-\frac{DL}{2}-1]\int  \dd^E \alpha \,
	\delta(1- \sum_{i=1}^E \alpha_i) \, \\ 
 & \times  \left[
     \cF v_1\cdot v_2 \widetilde M^{\text{adj}}_{j _1,j _2}+(D L-2 E+2) P_{j_1}\cdot v_1 P_{j_2}\cdot v_2\right] \,  
 \cF^{\frac{D L}{2}-E} \cU^{-\frac{D L}{2}-\frac{D}{2}+E-2} \,,
     \end{aligned}
\end{equation}
where $P_{j_r}^\mu=\sum_i \widetilde M^{\text{adj}}_{j_r i}
Q_i^\mu$. This agrees with the general 
expression in Ref.~\cite{Gluza:2010rn}. Notice that explicit factors of $\cF$ appear in the numerator. In the main text we consider numerators where the rank is not homogeneous. In those cases we combine them into a single polynomial where gamma factors and the exponents of $\cU$ and $\cF$ are those associated with the highest rank. 

\section{Local finiteness in momentum and in parameter space}
\label{appendix-local-finiteness}
In this Appendix we show that locally finite integrals in momentum space yield locally finite integrals in parameter space.

To make this statement precise, we first need to clarify the notion of local finiteness.
Intuitively, we can call an integral locally finite if there are no cancellations of divergences from different integration regions.
In other words, the integral converges on any measurable subset of the integration domain.
In particular, we intend the integral to be evaluatable when $D=4$.
This notion is equivalent to that of absolute convergence%
\footnote{Note that convergence theorems for Feynman integrals---such as Weinberg's theorem~\cite{Weinberg:1959nj,Zimmermann:1968mu} and BFP theorem---establish absolute convergence rather than local finiteness.}
or Lebesgue integrability.
Indeed, local finiteness implies in particular that the integral converges on the subset of the integration domain where the integrand is positive, so that absolute convergence follows; conversely, absolute convergence puts a finite upper bound on the integral value on any measurable subset.

Let us now consider a Feynman integral in momentum space~\eqref{integral-loop-withnumerator} and suppose that it is locally finite (absolutely convergent) in \(D=4\).
Introducing the Feynman parameters, we can write:
\newcommand{\ddell}{\dd \ellset}
\newcommand{\ddfeyn}{\dd \feynpset}
\begin{equation}
  \int \ddell \;
  \frac{\Numer(\ellset,\kset)}{\Den_1 \cdots \Den_E}
  = \int \ddell \int \ddfeyn \;
  \frac{\Numer(\ellset,\kset)}{\left(\alpha_1 \Den_1 + \cdots \alpha_E \Den_E \right)^E}\, ,
  \label{integral-loop-mixed}
\end{equation}
with the integration measure defined as \(\ddell \equiv \prod_{j=1}^L \dd^4 \ell_j\) and \(\ddfeyn \equiv \Gamma(E) \, \dd^E \alpha \, \delta \left( 1 - \sum_e \alpha_e \right)\).
At this point, the mixed-representation integral on the right-hand side of \Eqn{integral-loop-mixed} should be understood as an iterated integral: first over \(\alpha\)s, then over \(\ell\)s.
If we can show that this iterated integral converges absolutely, then absolute convergence of the parameter-space integral will follow automatically.
The reason is that, in an absolutely convergent mixed-representation integral we can safely exchange the integration order%
\footnote{In dimensional or analytic regularization this exchange is standard 
 but here we do not assume a regularization scheme.}
and integrate out the loop momenta.
The resulting Feynman-parametric integral is then a partially integrated absolutely convergent integral, which is guaranteed to converge absolutely as well.

Absolute convergence of the mixed-representation integral is straightforward to establish in the case of Euclidean denominators,
\begin{equation}
  \Den_e = q_{e,0}^2 + \pmb{q}_e^2 + m_e^2,
\end{equation}
where \(q_e = (q_{e,0}, \pmb{q}_e)\) is the corresponding edge momentum and \(m_e\) its mass.
Indeed, since all denominators are non-negative, we can apply Feynman's trick directly to the absolute value of the integrand:
\begin{equation}
  \begin{aligned}
    \int \ddell \int \ddfeyn
    \left\lvert \frac{\Numer(\ellset,\kset)}{\left(\alpha_1 \Den_1 + \cdots \alpha_E \Den_E \right)^E} \right\rvert
    &=
    \int \ddell \int \ddfeyn \;
    \frac{\left\lvert \Numer(\ellset,\kset) \right\rvert}{\left(\alpha_1 \Den_1 + \cdots \alpha_E \Den_E \right)^E}
    \\[1ex]
    &=
    \int \ddell \;
    \frac{\left\lvert \Numer(\ellset,\kset) \right\rvert}{\Den_1 \cdots \Den_E}
    =
    \int \ddell
    \left\lvert \frac{\Numer(\ellset,\kset)}{\Den_1 \cdots \Den_E} \right\rvert < \infty \, .
  \end{aligned}
\end{equation}
Unfortunately, this does not work for Minkowskian denominators,
\begin{equation}
  \Den_e = q_{e,0}^2 - \pmb{q}_e^2 - m_e^2 + i \varepsilon,
\end{equation}
due to possible cancellations between terms in the mixed denominator.
In fact, the momentum-space integral itself may not converge absolutely even if it satisfies the UV power-counting condition.
To see this, consider a bubble integral in two dimensions with a linear numerator,
\begin{equation}
  \int \dd^2 \ell \, \frac{\ell \cdot k}{\left( \ell^2 - m^2 + i \varepsilon \right) \left( \left( \ell - k \right)^2 - m^2 + i \varepsilon \right)} \, .
  \label{uv-div-bubble}
\end{equation}
This integral is UV-finite by power counting.
However, the same integral can be written in lightcone coordinates \(\ell_{\pm} = \ell_0 \pm \ell_1\) as,
\begin{equation}
  \int \dd \ell_+ \dd \ell_- \frac{\ell_+ k_- + \ell_- k_+}{\left( \ell_+ \ell_- - m^2 + i \varepsilon \right) \left( \left( \ell_+ - k_+ \right) \left( \ell_- - k_- \right) - m^2 + i \varepsilon \right)} \,.
\end{equation}
The \(\ell_+\) integration diverges logarithmically for almost any choice of \(\ell_-\), which means that the original integral~\eqref{uv-div-bubble} cannot converge absolutely.

This apparent breakdown of Weinberg's theorem is due to the fact that its original formulation~\cite{Weinberg:1959nj} assumes Euclidean denominators.
To restore the validity of the theorem in Minkowskian case, one can replace the standard \(i \varepsilon\) in the denominators with Zimmermann's prescription,
\begin{equation}
  \Den_{e, Z} = q_{e,0}^2 - \pmb{q}_e^2 - m_e^2 + i \varepsilon \left( \pmb{q}_e^2 + m_e^2 \right),
  \label{Zimmermann-prescription}
\end{equation}
which was introduced in Ref.~\cite{Zimmermann:1968mu} and used later to prove absolute convergence of integrals in the context of BPHZ renormalization~\cite{Zimmermann:1969jj}.
The merit of Zimmermann's prescription is that it puts a Euclidean 
upper and lower bound on the integrand, so that the convergence of 
Minkowskian integrals becomes equivalent to that 
of Euclidean ones.
As a consequence, absolutely convergent momentum-space integrals lead to absolutely convergent parameter-space integrals.
Furthermore, one can prove that the standard parameter-space integral (with the usual \(i \varepsilon\) prescription) converges absolutely as well, and all integrals agree in the~\(\varepsilon \to 0\) limit~\cite{Zimmermann:1968mu,Lowenstein:1975ku}.
Therefore, we can formulate the following statement on absolute convergence in momentum and parameter space which we expect to hold in general assuming that we drop regularization:
\begin{quote}
  Suppose that a four-dimensional momentum-space Feynman integral is IR-finite and satisfies Weinberg's UV power-counting criterion.
  Then it is absolutely convergent as long as Zimmermann's \(i \varepsilon\) prescription~\eqref{Zimmermann-prescription} is used.
  The corresponding parameter-space integral with the standard Feynman \(i \varepsilon\) prescription is absolutely convergent as well, and the two integrals have the same value in the \(\varepsilon \to 0\) limit.
\end{quote}

\section{BFP convergence with several monomials in the numerator}
\label{app:several-mons}

The BFP theorem gives a convergence condition for a parameter-space Feynman integral in the special case when the numerator is a lone monomial.
In this Appendix we investigate the general case with the numerator being a linear combination of monomials.
We will argue that no local cancellations of divergences between different numerator monomials are possible.

More precisely, consider an integral of the form
\begin{equation}
  I = \int_{\mathbb{R}_{>0}^h} \frac{\dd \feynp_1}{\feynp_1} \cdots 
  \frac{\dd \feynp_h}{\feynp_h} \, 
  \frac{f (\feynpset)}{g (\feynpset)} \,,
  \label{euler-mellin}
\end{equation}
where the numerator is a polynomial,
\begin{equation}
  f (\feynpset) = \sum_i c_i \, \feynpset^{\nvec_i} \,,
  \label{general-numerator}
\end{equation}
with \(c_i \neq 0\),
and the denominator~\(g (\feynpset)\) is a completely non-vanishing polynomial raised to some complex power~\(n_g\) with a positive real part, \(\operatorname{Re} n_g > 0\).
The imaginary part of~\(n_g\) does not affect the absolute value of \(g (\feynpset)\), so for the purposes of studying absolute convergence we can take it to be zero.
Thus we have \(n_g \in \mathbb{R}_{>0}\) and the Newton polytope \(\Newton(g)\) is well-defined.

The BFP theorem teaches us that for a numerator consisting of a single monomial, \(f (\feynpset) = c \, \feynpset^{\nvec}\), the integral~\eqref{euler-mellin} converges absolutely as long as \(\nvec \in \operatorname{int} (\Newton (g))\).
We will argue that this result extends to the case of a general numerator~\eqref{general-numerator}: namely, the integral~\eqref{euler-mellin} converges absolutely if and only if \(\nvec_i \in \inter (\Newton (g))\) for all \(i\), or equivalently, \(\Newton (f) \subset \inter (\Newton (g))\).%
\footnote{As explained in section~\ref{polytopes}, \(\inter \Newton(g) = \relint \Newton(g)\) for a full dimensional polytope \(\Newton (g)\).
If the polytope is less than full dimensional, then its interior is empty, and the integral diverges.}
In other words, each numerator monomial yields an absolutely convergent integral separately, and no cancellations between different terms are possible.

Sufficiency of this condition for absolute convergence of the integral follows trivially from the BFP theorem.
To show that it is also necessary, let us assume the opposite: \(\Newton (f) \not \subset \inter \Newton (g)\).
There are two possibilities: either some part of \(\Newton (f)\) lies strictly outside \(\Newton (g)\) (as in Fig.~\ref{fig:div-newt-outside}), or \(\Newton (f)\) lies inside \(\Newton (g)\) and touches its boundary (as in Fig.~\ref{fig:div-newt-boundary}).
We start with the former case.

\newcommand{\wvec}{\pmb{w}}

\begin{figure}[t]
  \newcommand{\common}{
    \draw[step=1, help lines] (-0.9,-0.9) grid (3.9,3.9);
    \begin{scope}
      [->]
      \draw (-1,0) -- (4,0);
      \draw (0,-1) -- (0,4);
    \end{scope}
    \draw[thick, fill=\facetcolor, fill opacity=0.6] (0,0) -- (3,0) -- (0,3) -- cycle;
    \begin{scope}
      \fill (0,0) circle (2pt);
      \fill (3,0) circle (2pt);
      \fill (0,3) circle (2pt);
    \end{scope}
  }
  \centering
  \subfloat[]{%
    \label{fig:div-newt-outside}
    \begin{tikzpicture}
      [x=5ex, y=5ex]
      \common
      \draw[pattern={Lines[angle=70, distance=4pt, line width=0.3pt]}] (2,2) -- (2,0) -- (0,1) -- cycle;
      \begin{scope}
        \fill (2,2) circle (2pt);
        \fill (2,0) circle (2pt);
        \fill (0,1) circle (2pt);
      \end{scope}
      \draw[dashed] (2.5,1.5) -- (1.5,2.5);
      \draw[->] (2,2) -- +(0.5,0.5) node[above] {\(\wvec\)};
    \end{tikzpicture}
  }
  \qquad
  \subfloat[]{%
    \label{fig:div-newt-boundary}
    \begin{tikzpicture}
      [x=5ex, y=5ex]
      \common
      \draw[pattern={Lines[angle=70, distance=4pt, line width=0.3pt]}] (2,1) -- (1,2) -- (1,1) -- cycle;
      \begin{scope}
        \fill (2,1) circle (2pt);
        \fill (1,2) circle (2pt);
        \fill (1,1) circle (2pt);
      \end{scope}
      \draw[->] (1.5,1.5) -- +(0.5,0.5) node[above] {\(\wvec\)};
    \end{tikzpicture}
  }
  \caption{%
    Two possible configurations giving rise to a divergent parameter-space integral.
    The shaded region depicts the Newton polytope of the denominator, \(\Newton (g)\).
    The hatched region depicts the Newton polytope of the numerator, \(\Newton (f)\). 
  }
  \label{fig:div-newt}
\end{figure}
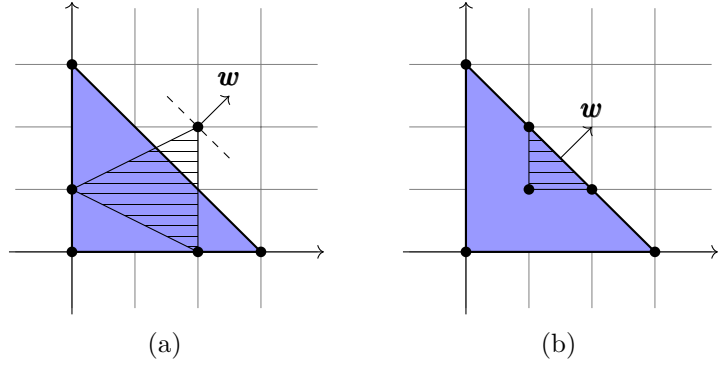

If \(\Newton (f) \setminus \Newton (g) \neq \emptyset\), then there exists a vertex \(\nvec_k \in \Newton (f)\) [in the $V$-representation] such that \(\nvec_k \notin \Newton (g)\).
Correspondingly, there exists a scaling vector~\(\wvec\) such that \(\wvec \cdot \nvec_k > \wvec \cdot \nvec\), where \(\nvec\) is any other vertex in \(\Newton (f)\) or \(\Newton (g)\) (see Fig.~\ref{fig:div-newt-outside}).
Therefore, upon rescaling \(\feynpset \to \rho^{\wvec} \feynpset\) the numerator monomial \(c_k \, \feynpset^{\nvec_k}\) strictly dominates any other monomial in \(f\) and \(g\) in the limit \(\rho \to \infty\).
As a result, the integral cannot be absolutely convergent, because the one-dimensional integration over \(\rho\) diverges.
No cancellation is possible simply because only one monomial dominates in this regime.

To illustrate this behavior, consider the following example.
Let us take \(h=2\),
\begin{equation}
    g (\feynpset) = 1 + \alpha_1^3 + \alpha_2^3,
    \quad \textrm{and}\quad
    f (\feynpset) = c_1 \alpha_1^2 \alpha_2^2 + c_2 \alpha_1^2 + c_3 \alpha_2 \,.
\end{equation}
One possible choice for a scaling vector~\(\wvec\) yielding a divergent integration is \(\wvec = (1, 1)\) (see Fig.~\ref{fig:div-newt-outside}). 
Indeed, upon a logarithmic change of variables \(y_i = \log \alpha_i\) the integral becomes
\begin{equation}
    I = \int_{\mathbb{R}^2} \dd y_1 \dd y_2 \frac{c_1 e^{2(y_1 + y_2)} + c_2 e^{2y_1} + c_3 e^{y_2}}{1 + e^{3y_1} + e^{3y_2}}.
\end{equation}
Rotating the coordinate frame as \(\tilde{y}_1 = y_1 + y_2, \tilde{y}_2 = y_1 - y_2\), we can further rewrite it as
\begin{equation}
    I = \frac{1}{2} \int_{\mathbb{R}^2} \dd \tilde{y}_1 \dd \tilde{y}_2 \frac{c_1 e^{2\tilde{y}_1} + c_2 e^{\tilde{y}_1 + \tilde{y}_2} + c_3 e^{(\tilde{y}_1 - \tilde{y}_2)/2}}{1 + e^{3(\tilde{y}_1 + \tilde{y}_2)/2} + e^{3(\tilde{y}_1 - \tilde{y}_2)/2}} \,.
\end{equation}
In these new coordinates, the \(\tilde{y}_1\) axis is aligned with \(\wvec\), and \(\tilde{y}_1\) can be identified with \(\log \rho\).
Clearly, the integral over \(\tilde{y}_1\) diverges as \(\tilde{y}_1 \to + \infty\) for any choice of \(\tilde{y}_2\) as long as the ``exterior'' monomial \(\alpha_1^2 \alpha_2^2\) has a non-vanishing coefficient \(c_1 \neq 0\), because this term dominates any other term in the numerator and the denominator.

Let us now consider the second possibility, \(\Newton (f) \subset \Newton (g)\) and \(\Newton (f) \cap \partial \Newton (g) \neq \emptyset\), where \(\partial\) denotes the boundary.
In this case, a non-empty subset of exponents \(\set{\nvec_k \vert k \in K} \subset \Newton (f)\) lies on some facet of \(\Newton (g)\).
Taking \(\wvec\) to be the normal vector of this facet, we find that upon rescaling \(\feynpset \to \rho^{\wvec} \feynpset\) the corresponding numerator monomials \(\set{\feynpset^{\nvec_k}}\) become dominant together with the denominator monomials whose exponents lie on the same facet.
As a result, generically we get a logarithmic divergence along \(\wvec\) as \(\rho \to \infty\).

In general, there may be more than one monomial in \(\set{\feynpset^{\nvec_k}}\), in which case one might wonder if term-by-term divergences cancel out for some choice of the coefficients \(c_k\).
We argue that this cannot happen because different monomials have different functional dependence on the integration variables which factors out of the divergent integral along \(\wvec\).
To see this, consider an example with,
\begin{equation}
    g (\feynpset) = 1 + \alpha_1^3 + \alpha_2^3,
    \quad
    f (\feynpset) = c_1 \alpha_1^2 \alpha_2 + c_2 \alpha_1 \alpha_2^2 + c_3 \alpha_1 \alpha_2
\end{equation}
(see Fig.~\ref{fig:div-newt-boundary}).
Performing the same change of variables as in our previous example, we can rewrite the integral as,
\begin{equation}
    I = \frac{1}{2} \int_{\mathbb{R}^2} \dd \tilde{y}_1 \dd \tilde{y}_2
    \frac{c_1 e^{(3 \tilde{y}_1 + \tilde{y}_2)/2} + c_2 e^{(3 \tilde{y}_1 - \tilde{y}_2)/2} + c_3 e^{\tilde{y}_1}}{1 + e^{3(\tilde{y}_1 + \tilde{y}_2)/2} + e^{3(\tilde{y}_1 - \tilde{y}_2)/2}} \,.
\end{equation}
The first two terms of the numerator yield divergent integrals over \(\tilde{y}_1\) as \(\tilde{y}_1 \to + \infty\).
Note that dependence on the remaining integration variable \(\tilde{y}_2\) factors out of the divergent integral together with the monomial coefficients:
\begin{equation}
\begin{aligned}
    \int \dd &\tilde{y}_1 \frac{c_1 e^{(3 \tilde{y}_1 + \tilde{y}_2)/2} + c_2 e^{(3 \tilde{y}_1 - \tilde{y}_2)/2}}{1 + e^{3(\tilde{y}_1 + \tilde{y}_2)/2} + e^{3(\tilde{y}_1 - \tilde{y}_2)/2}}
    = \\&\hspace*{15mm}
    \left( c_1 e^{\tilde{y}_2 / 2} + c_2 e^{-\tilde{y}_2 / 2} \right)
    \int \dd \tilde{y}_1 \frac{e^{3 \tilde{y}_1 / 2}}{1 + e^{3(\tilde{y}_1 + \tilde{y}_2)/2} + e^{3(\tilde{y}_1 - \tilde{y}_2)/2}} \,.
\end{aligned}
\end{equation}
The integral therefore diverges for almost any value of 
\(\tilde{y}_2\) independent of the coefficients, 
and hence the original integral cannot converge absolutely.

Divergences from different numerator terms cannot cancel each 
other yielding a locally finite integral, that is an 
absolutely convergent one.  Within dimensional regularization,
the integral may still turn be finite due to a non-local 
cancellation of divergences.

\bibliographystyle{apsrev4-2}
\bibliography{tropical.bib}

\end{document}

%% file: faces_Example.tex
\begin{tikzpicture}%
	[scale=1.000000,
	back/.style={loosely dotted, thin},
	edge/.style={color=black, dashed, thick},
	facet/.style={fill=\facetcolor, fill opacity=0.400000},
	vertex/.style={inner sep=1pt,circle,draw=black,fill=black,thick}]
	%
	%
	\draw [help lines, very thin] (0,0) grid (3,3);
	\coordinate (a) at (-0.2,0);
	\coordinate (a1) at (0,-0.2);
	\coordinate (b) at (3.3,0);
	\coordinate (b1) at (0,3.3);
	\draw[->, very thin] (a) -- (b);
	\draw[->, very thin] (a1) -- (b1);
	\coordinate (0.00000, 1.00000) at (0.00000, 1.00000);
	\coordinate (0.00000, 3.00000) at (0.00000, 3.00000);
	\coordinate (1.00000, 0.00000) at (1.00000, 0.00000);
	\coordinate (3.00000, 0.00000) at (3.00000, 0.00000);
	\fill[facet] (3.00000, 0.00000) -- (0.00000, 3.00000) -- (0.00000, 1.00000) -- (1.00000, 0.00000) -- cycle {};
	\draw[edge] (0.00000, 1.00000) -- (0.00000, 3.00000);
	\draw[edge] (0.00000, 1.00000) -- (1.00000, 0.00000);
	\draw[edge] (0.00000, 3.00000) -- (3.00000, 0.00000);
	\draw[edge] (1.00000, 0.00000) -- (3.00000, 0.00000);
	
	\node[vertex] at (0.00000, 1.00000)     {};
	\node[vertex] at (0.00000, 3.00000)     {};
	\node[vertex] at (1.00000, 0.00000)     {};
	\node[vertex] at (3.00000, 0.00000)     {};
	
\end{tikzpicture}

%% file: triangle_poly1.tex
\begin{tikzpicture}%
	[scale=1.000000,
	back/.style={loosely dotted, thin},
	edge/.style={color=black, thick},
	facet/.style={fill=\facetcolor,fill opacity=0.400000},
	vertex/.style={inner sep=1pt,circle,draw=black,fill=black,thick}]
%
%
  \draw [help lines] (0,0) grid (3,3);
  \coordinate (a) at (-0.2,0);
\coordinate (a1) at (0,-0.2);
  \coordinate (b) at (3.3,0);
    \coordinate (b1) at (0,3.3);
  \draw[->] (a) -- (b);
          \draw[->] (a1) -- (b1);

\coordinate (0.00000, 1.00000) at (0.00000, 1.00000);
\coordinate (0.00000, 2.00000) at (0.00000, 2.00000);
\coordinate (1.00000, 0.00000) at (1.00000, 0.00000);
\coordinate (2.00000, 0.00000) at (2.00000, 0.00000);
\fill[facet] (2.00000, 0.00000) -- (0.00000, 2.00000) -- (0.00000, 1.00000) -- (1.00000, 0.00000) -- cycle {};
\draw[edge] (0.00000, 1.00000) -- (0.00000, 2.00000);
\draw[edge] (0.00000, 1.00000) -- (1.00000, 0.00000);
\draw[edge] (0.00000, 2.00000) -- (2.00000, 0.00000);
\draw[edge] (1.00000, 0.00000) -- (2.00000, 0.00000);
\node[vertex] at (0.00000, 1.00000)     {};
\node[vertex] at (0.00000, 2.00000)     {};
\node[vertex] at (1.00000, 0.00000)     {};
\node[vertex] at (2.00000, 0.00000)     {};
\end{tikzpicture}

%% file: triangle_poly2.tex
\begin{tikzpicture}%
	[scale=1.000000,
	back/.style={loosely dotted, thin},
	edge/.style={color=black, thick},
	facet/.style={fill=\facetcolor,fill opacity=0.400000},
	vertex/.style={inner sep=1pt,circle,draw=black,fill=black,thick}]
%
  \draw [help lines] (0,0) grid (3,3);
  \coordinate (a) at (-0.2,0);
\coordinate (a1) at (0,-0.2);
  \coordinate (b) at (3.3,0);
    \coordinate (b1) at (0,3.3);
  \draw[->] (a) -- (b);
    \draw[->] (a1) -- (b1);

\coordinate (0.00000, 1.00000) at (0.00000, 1.00000);
\coordinate (0.00000, 3.00000) at (0.00000, 3.00000);
\coordinate (1.00000, 0.00000) at (1.00000, 0.00000);
\coordinate (3.00000, 0.00000) at (3.00000, 0.00000);
\fill[facet] (3.00000, 0.00000) -- (0.00000, 3.00000) -- (0.00000, 1.00000) -- (1.00000, 0.00000) -- cycle {};
\draw[edge] (0.00000, 1.00000) -- (0.00000, 3.00000);
\draw[edge] (0.00000, 1.00000) -- (1.00000, 0.00000);
\draw[edge] (0.00000, 3.00000) -- (3.00000, 0.00000);
\draw[edge] (1.00000, 0.00000) -- (3.00000, 0.00000);

\node[vertex] at (0.00000, 1.00000)     {};
\node[vertex] at (0.00000, 3.00000)     {};
\node[vertex] at (1.00000, 0.00000)     {};
\node[vertex] at (3.00000, 0.00000)     {};
\node[vertex] at (1.00000, 1.00000)     {};

\end{tikzpicture}

%% file: triangle_poly3.tex
\begin{tikzpicture}%
	[scale=1.000000,
	back/.style={loosely dotted, thin},
	edge/.style={color=black, thick},
	facet/.style={fill=\facetcolor,fill opacity=0.400000},
	vertex/.style={inner sep=1pt,circle,draw=black,fill=black,thick}]
%
%
  \draw [help lines] (0,0) grid (4,4);
  \coordinate (a) at (-0.2,0);
\coordinate (a1) at (0,-0.2);
  \coordinate (b) at (4.3,0);
    \coordinate (b1) at (0,4.3);
  \draw[->] (a) -- (b);
    \draw[->] (a1) -- (b1);

\coordinate (0.00000, 1.00000) at (0.00000, 1.00000);
\coordinate (0.00000, 4.00000) at (0.00000, 4.00000);
\coordinate (1.00000, 0.00000) at (1.00000, 0.00000);
\coordinate (4.00000, 0.00000) at (4.00000, 0.00000);
\fill[facet] (4.00000, 0.00000) -- (0.00000, 4.00000) -- (0.00000, 1.00000) -- (1.00000, 0.00000) -- cycle {};
\draw[edge] (0.00000, 1.00000) -- (0.00000, 4.00000);
\draw[edge] (0.00000, 1.00000) -- (1.00000, 0.00000);
\draw[edge] (0.00000, 4.00000) -- (4.00000, 0.00000);
\draw[edge] (1.00000, 0.00000) -- (4.00000, 0.00000);
\node[vertex] at (0.00000, 1.00000)     {};
\node[vertex] at (0.00000, 4.00000)     {};
\node[vertex] at (1.00000, 0.00000)     {};
\node[vertex] at (4.00000, 0.00000)     {};

\node[vertex] at (1.00000, 1.00000)     {};
\node[vertex] at (1.00000, 2.00000)     {};
\node[vertex] at (2.00000, 1.00000)     {};

\end{tikzpicture}

%% file: rank1-box.tex
\begin{tikzpicture}%
	[x={(0.626699cm, -0.082367cm)},
	y={(0.779261cm, 0.066206cm)},
	z={(0.000028cm, 0.994401cm)},
	scale=1.5,
	back/.style={loosely dotted, thin},
	edge/.style={color=black, thick},
	facet/.style={fill=\facetcolor,fill opacity=0.6},
	vertex/.style={inner sep=1pt,circle,draw=black!25!black,fill=black!75!black,thick}]
%
%
\coordinate (0.00000, 2.00000, 0.00000) at (0.00000, 2.00000, 0.00000);
\coordinate (0.00000, 2.00000, 1.00000) at (0.00000, 2.00000, 1.00000);
\coordinate (0.00000, 3.00000, 0.00000) at (0.00000, 3.00000, 0.00000);
\coordinate (1.00000, 2.00000, 0.00000) at (1.00000, 2.00000, 0.00000);
\coordinate (2.00000, 0.00000, 2.00000) at (2.00000, 0.00000, 2.00000);
\coordinate (2.00000, 0.00000, 3.00000) at (2.00000, 0.00000, 3.00000);
\coordinate (2.00000, 1.00000, 2.00000) at (2.00000, 1.00000, 2.00000);
\coordinate (3.00000, 0.00000, 2.00000) at (3.00000, 0.00000, 2.00000);
\draw[edge,back] (0.00000, 2.00000, 0.00000) -- (0.00000, 2.00000, 1.00000);
\draw[edge,back] (0.00000, 2.00000, 0.00000) -- (0.00000, 3.00000, 0.00000);
\draw[edge,back] (0.00000, 2.00000, 1.00000) -- (0.00000, 3.00000, 0.00000);
\draw[edge,back] (0.00000, 2.00000, 1.00000) -- (2.00000, 0.00000, 3.00000);
\node[vertex] at (0.00000, 2.00000, 1.00000)     {};
\fill[facet] (3.00000, 0.00000, 2.00000) -- (1.00000, 2.00000, 0.00000) -- (0.00000, 2.00000, 0.00000) -- (2.00000, 0.00000, 2.00000) -- cycle {};
\fill[facet] (3.00000, 0.00000, 2.00000) -- (1.00000, 2.00000, 0.00000) -- (0.00000, 3.00000, 0.00000) -- (2.00000, 1.00000, 2.00000) -- cycle {};
\fill[facet] (3.00000, 0.00000, 2.00000) -- (2.00000, 0.00000, 2.00000) -- (2.00000, 0.00000, 3.00000) -- cycle {};
\fill[facet] (3.00000, 0.00000, 2.00000) -- (2.00000, 0.00000, 3.00000) -- (2.00000, 1.00000, 2.00000) -- cycle {};
\draw[edge] (0.00000, 2.00000, 0.00000) -- (1.00000, 2.00000, 0.00000);
\draw[edge] (0.00000, 2.00000, 0.00000) -- (2.00000, 0.00000, 2.00000);
\draw[edge] (0.00000, 3.00000, 0.00000) -- (1.00000, 2.00000, 0.00000);
\draw[edge] (0.00000, 3.00000, 0.00000) -- (2.00000, 1.00000, 2.00000);
\draw[edge] (1.00000, 2.00000, 0.00000) -- (3.00000, 0.00000, 2.00000);
\draw[edge] (2.00000, 0.00000, 2.00000) -- (2.00000, 0.00000, 3.00000);
\draw[edge] (2.00000, 0.00000, 2.00000) -- (3.00000, 0.00000, 2.00000);
\draw[edge] (2.00000, 0.00000, 3.00000) -- (2.00000, 1.00000, 2.00000);
\draw[edge] (2.00000, 0.00000, 3.00000) -- (3.00000, 0.00000, 2.00000);
\draw[edge] (2.00000, 1.00000, 2.00000) -- (3.00000, 0.00000, 2.00000);
\node[vertex] at (0.00000, 2.00000, 0.00000)     {};
\node[vertex] at (0.00000, 3.00000, 0.00000)     {};
\node[vertex] at (1.00000, 2.00000, 0.00000)     {};
\node[vertex] at (2.00000, 0.00000, 2.00000)     {};
\node[vertex] at (2.00000, 0.00000, 3.00000)     {};
\node[vertex] at (2.00000, 1.00000, 2.00000)     {};
\node[vertex] at (3.00000, 0.00000, 2.00000)     {};
\end{tikzpicture}

%% file: rank2-box.tex
\begin{tikzpicture}%
	[x={(0.626699cm, -0.082367cm)},
	y={(0.779261cm, 0.066206cm)},
	z={(0.000028cm, 0.994401cm)},
	scale=1.5,
	back/.style={loosely dotted, thin},
	edge/.style={color=black, thick},
	facet/.style={fill=\facetcolor,fill opacity=0.6},
	vertex/.style={inner sep=1pt,circle,draw=black!25!black,fill=black!75!black,thick}]
%
%
\coordinate (0.00000, 2.00000, 0.00000) at (0.00000, 2.00000, 0.00000);
\coordinate (0.00000, 2.00000, 2.00000) at (0.00000, 2.00000, 2.00000);
\coordinate (0.00000, 4.00000, 0.00000) at (0.00000, 4.00000, 0.00000);
\coordinate (2.00000, 0.00000, 2.00000) at (2.00000, 0.00000, 2.00000);
\coordinate (2.00000, 0.00000, 4.00000) at (2.00000, 0.00000, 4.00000);
\coordinate (2.00000, 2.00000, 0.00000) at (2.00000, 2.00000, 0.00000);
\coordinate (2.00000, 2.00000, 2.00000) at (2.00000, 2.00000, 2.00000);
\coordinate (4.00000, 0.00000, 2.00000) at (4.00000, 0.00000, 2.00000);
\draw[edge,back] (0.00000, 2.00000, 0.00000) -- (0.00000, 2.00000, 2.00000);
\draw[edge,back] (0.00000, 2.00000, 0.00000) -- (0.00000, 4.00000, 0.00000);
\draw[edge,back] (0.00000, 2.00000, 2.00000) -- (0.00000, 4.00000, 0.00000);
\draw[edge,back] (0.00000, 2.00000, 2.00000) -- (2.00000, 0.00000, 4.00000);
\node[vertex] at (0.00000, 2.00000, 2.00000)     {};
\fill[facet] (4.00000, 0.00000, 2.00000) -- (2.00000, 2.00000, 0.00000) -- (0.00000, 4.00000, 0.00000) -- (2.00000, 2.00000, 2.00000) -- cycle {};
\fill[facet] (4.00000, 0.00000, 2.00000) -- (2.00000, 0.00000, 2.00000) -- (0.00000, 2.00000, 0.00000) -- (2.00000, 2.00000, 0.00000) -- cycle {};
\fill[facet] (4.00000, 0.00000, 2.00000) -- (2.00000, 0.00000, 4.00000) -- (2.00000, 2.00000, 2.00000) -- cycle {};
\fill[facet] (4.00000, 0.00000, 2.00000) -- (2.00000, 0.00000, 2.00000) -- (2.00000, 0.00000, 4.00000) -- cycle {};
\draw[edge] (0.00000, 2.00000, 0.00000) -- (2.00000, 0.00000, 2.00000);
\draw[edge] (0.00000, 2.00000, 0.00000) -- (2.00000, 2.00000, 0.00000);
\draw[edge] (0.00000, 4.00000, 0.00000) -- (2.00000, 2.00000, 0.00000);
\draw[edge] (0.00000, 4.00000, 0.00000) -- (2.00000, 2.00000, 2.00000);
\draw[edge] (2.00000, 0.00000, 2.00000) -- (2.00000, 0.00000, 4.00000);
\draw[edge] (2.00000, 0.00000, 2.00000) -- (4.00000, 0.00000, 2.00000);
\draw[edge] (2.00000, 0.00000, 4.00000) -- (2.00000, 2.00000, 2.00000);
\draw[edge] (2.00000, 0.00000, 4.00000) -- (4.00000, 0.00000, 2.00000);
\draw[edge] (2.00000, 2.00000, 0.00000) -- (4.00000, 0.00000, 2.00000);
\draw[edge] (2.00000, 2.00000, 2.00000) -- (4.00000, 0.00000, 2.00000);
\node[vertex] at (0.00000, 2.00000, 0.00000)     {};
\node[vertex] at (0.00000, 4.00000, 0.00000)     {};
\node[vertex] at (2.00000, 0.00000, 2.00000)     {};
\node[vertex] at (2.00000, 0.00000, 4.00000)     {};
\node[vertex] at (2.00000, 2.00000, 0.00000)     {};
\node[vertex] at (2.00000, 2.00000, 2.00000)     {};
\node[vertex] at (4.00000, 0.00000, 2.00000)     {};
\node[blue] at (1.00000, 2.00000, 1.00000)     {$*$};
\node[blue] at (2.00000, 1.00000, 2.00000)     {$*$};
\end{tikzpicture}